\documentclass[aps,prb,showpacs,twocolumn,superscriptaddress]{revtex4-1}
\usepackage{bm,color,amsmath,amssymb,mathrsfs,latexsym,graphicx,psfrag,accents,float}
\pdfoutput=1

\newcommand{\ins}[1]{\;\;\;\;\text{#1}\;\;\;\;}

\newcommand{\partdif}[2]{\frac{ \partial #1}{\partial #2}}

\newcommand{\kpar}{\boldsymbol{k}_{\parallel}}

\newcommand{\gam}[1]{\Gamma_{#1}}

\newcommand{\cfv}{C_{4v}}

\newcommand{\cnv}{C_{nv}}

\newcommand{\cala}{{\cal A}}
\newcommand{\calb}{{\cal B}}
\newcommand{\calc}{{\cal C}}
\newcommand{\cald}{{\cal D}}

\newcommand{\calf}{{\cal F}}
\newcommand{\calg}{{\cal G}}
\newcommand{\calj}{{\cal J}}
\newcommand{\calk}{{\cal K}}
\newcommand{\call}{{\cal L}}
\newcommand{\caln}{{\cal N}}
\newcommand{\calo}{{\cal O}}
\newcommand{\calp}{{\cal P}}
\newcommand{\calr}{{\cal R}}
\newcommand{\cals}{{\cal S}}
\newcommand{\calt}{{\cal T}}
\newcommand{\calu}{{\cal U}}
\newcommand{\calv}{{\cal V}}

\newcommand{\cnt}{C_n {\cal T}}

\newcommand{\ct}{C_3}
\newcommand{\cst}{C_6{\cal T}}
\newcommand{\cft}{C_4{\cal T}}

\newcommand{\noi}[1]{\noindent (#1)}
\newcommand{\imp}{\;\;\Rightarrow\;\;}
\newcommand{\mo}{\text{-}1}
\newcommand{\minus}{\text{-}}
\newcommand{\oneover}[1]{\tfrac{1}{#1}}

\newcommand{\braket}[2]{\big\langle #1 \big| #2 \big\rangle}
\newcommand{\bra}[1]{\big\langle#1\big|}
\newcommand{\ket}[1]{\big|#1\big\rangle}

\newcommand{\half}{\tfrac{1}{2} }

\newcommand{\bea}{\begin{eqnarray}}
\newcommand{\enea}{\end{eqnarray}}
\newcommand{\beq}{\begin{equation}}
\newcommand{\eneq}{\end{equation}}

\newcommand{\pdg}[1]{{#1}^{\phantom{\dagger}}}

\newcommand{\eq}{=&\;}
\newcommand{\ab}{\alpha\beta}

\newcommand{\low}{L$\ddot{\text{o}}$wdin\;}

\newcommand{\W}{{\cal W}}

\newcommand{\bpm}{\begin{pmatrix}}
\newcommand{\epm}{\end{pmatrix}}

\newcommand{\bal}{\begin{align}}
\newcommand{\eal}{\end{align}}

\newcommand{\R}{\mathbb{R}}

\newcommand{\si}{\;\text{sin}\,}

\newcommand{\co}{\;\text{\text{cos}}\,}

\newcommand{\dg}[1]{#1^{\scriptstyle{\dagger}}}

\newcommand{\sma}[1]{\scriptscriptstyle{#1}}

\newcommand{\noc}{n_{\sma{{o}}}}

\newcommand{\Z}{\mathbb{Z}}

\newcommand{\qed}{\nobreak \ifvmode \relax \else
      \ifdim\lastskip<1.5em \hskip-\lastskip
      \hskip1.5em plus0em minus0.5em \fi \nobreak
      \vrule height0.75em width0.5em depth0.25em\fi}

\newcommand{\br}{\boldsymbol{r}}

\newcommand{\bG}{\boldsymbol{G}}

\begin{document}
\title{Berry-Phase Description of Topological Crystalline Insulators} 
 \author{A. Alexandradinata} \affiliation{Department of Physics, Princeton University, Princeton,
  NJ 08544} 
  \author{B. Andrei Bernevig} \affiliation{Department of Physics, Princeton University, Princeton,
  NJ 08544}
  

\begin{abstract}
We study a class of translational-invariant insulators with discrete rotational symmetry. These insulators have no spin-orbit coupling, and in some cases have no time-reversal symmetry as well, i.e., the relevant symmetries are purely crystalline. Nevertheless, topological phases exist which are distinguished by their robust surface modes. Like many well-known topological phases, their band topology is unveiled by the crystalline analog of Berry phases, i.e., parallel transport across certain non-contractible loops in the Brillouin zone. We also identify certain topological phases without any robust surface modes --  they are uniquely distinguished by parallel transport along \emph{bent} loops, whose shapes are determined by the symmetry group. Our findings have experimental implications in cold-atom systems, where the crystalline Berry phase has been directly measured.    
\end{abstract}
\date{\today}


\maketitle

For many well-known topological insulators (TI's), the existence of boundary bands is in one-to-one correspondence with the topology of the bulk wavefunctions, and robust boundary modes indicate a nontrivial phase\cite{fidkowski2011}. Examples include the Chern insulator and the quantum spin Hall insulator; these non-interacting insulators fall under ten well-known symmetry classes which are distinguished by time-reversal, particle-hole and chiral symmetries\cite{schnyder2009}. Given the completeness of this classification, attention has shifted to identifying topological phases which rely on other symmetries. The symmetries which are ubiquitous in condensed matter are the crystal space groups; among them the point groups involve transformations that preserve a spatial point. In a recent study of spin-orbit-free insulators with point groups, we identified the $\cnv$ groups as being able to host robust surface modes, for $n=3,4$ and $6$\cite{AAchen}. Here, `spin-orbit-free' describes both electronic phases with negligibly weak spin-orbit coupling, and intrinsically spinless systems such as photonic crystals. These are the first-known 3D TI with robust surface modes that are protected only by point groups, i.e., not requiring TRS, and not requiring spin -- the relevant symmetries are purely crystalline. In a related study\cite{fu2011}, it has also been shown that spin-orbit-free topological phases exist if the $C_n$ point group is combined with TRS, for $n=4$ and $6$. Henceforth, we refer to this combined group as $C_n+T$, where $n \in \{4,6\}$; for $\cnv$, it is understood that $n \in \{3,4,6\}$ only. To date, all experimentally-realized TI's are spin-orbit coupled; it is hoped that our study of spin-orbit-free phases enlarges the range of TI candidates.\\

\begin{figure}
\centering
\includegraphics[width=8.6 cm]{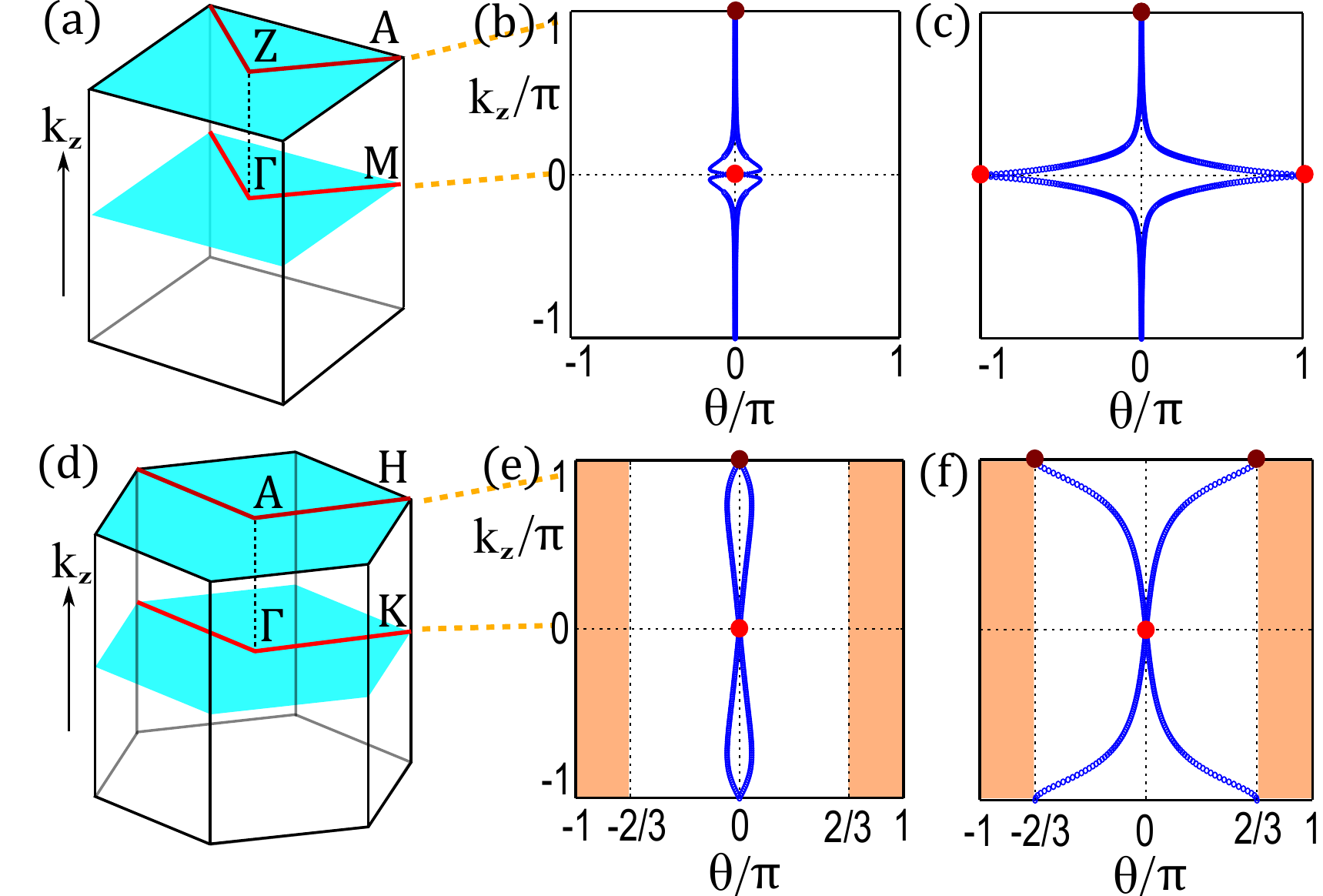}
\caption{(a) (resp.\ (d)) illustrates certain bent loops in the 3D Brillouin zone of a simple tetragonal (resp.\ hexagonal) lattice. (b) and (c): Berry-phase spectrum of trivial and strong phases respectively, for the model of (\ref{cfvmodel}). (e) and (f): Berry-phase spectrum of trivial and strong phases, for a $C_6+T$ model described in App.\ \ref{app:modelC6T}.}\label{fig:BZsWloops}
\end{figure} 

Our case study of the $C_n+T$ insulator is motivated by two issues: (i) In Ref. \onlinecite{fu2011}, we learned there exists two distinct gapped phases with $C_n+T$ symmetry, which are distinguished by robust surface modes; we refer to one as the trivial phase and the other a strong topological phase. It is unclear if these two phases are physically distinguishable if we experimentally probe the bulk instead of the surface. (ii) In this work we highlight the existence of a third topological phase which does not manifest robust surface modes. Does this `weak phase' have any physical consequence? One answer to (i) and (ii) may be found through holonomy, i.e., parallel transport along certain non-contractible loops in the Brillouin zone (BZ)\cite{leone2011,ekert2000,recati2002}.  A particle transported around a loop acquires a Berry-Zak phase\cite{berry1984,zak1982,zak1989}, which has recently been measured by Ramsey interference in cold-atom experiments\cite{atala2013,Tracy}. For the purpose of unveiling the bulk topology of all three phases, we find that not all non-contractible loops work. Straight loops are commonly studied in the geometric theory of polarization, due to their relation with Wannier functions\cite{kingsmith1993,Maryam2014}; however, these loops cannot identify our weak phase. Instead, we propose that all three phases are distinguished by parallel transport along \emph{bent} loops, whose shapes are determined by the symmetry group -- they are illustrated in Fig.\ \ref{fig:BZsWloops}(a) and (d), for the $C_4+T$ and $C_6+T$ groups respectively. Our formulation through holonomy simultaneously reveals a geometric connection with the theory of matrices in $SO(2m)$ -- we show that different sectors of ground states have a one-to-one correspondence with distinct classes of rotations.\\

Our second case study is the $\cnv$ insulator, which has a richer variety of gapped phases. For example, the $\cfv$ insulator is specified by two integer invariants called \emph{halved}-mirror chiralities ($\chi$)\cite{AAchen}; insulators with distinct $\chi$ are distinguished by surface bands with a unique $\Z$ topology. In this paper we explicitly formulate the halved chirality in terms of Berry phases, which allows for efficient computation of these invariants. In gapped phases with nonzero $\chi$, the Berry phases exhibit spectral flow, i.e., they interpolate across their maximally-allowed range as we tune a BZ parameter. Spectral flow is a unifying trait shared by many TI's, including the Chern insulator\cite{Haldane1988,thouless1982,qi2006,ProdanJMP2009}, the quantum spin Hall insulator\cite{yu2011,soluyanov2011}, and the inversion-symmetric TI with relative winding\cite{AA2014}. \\

The organization of our paper: in Sec.\ \ref{app:rvtightbinding}, we review how symmetries constrain the tight-binding Hamiltonian, which we employ throughout the paper. Our two case studies are presented in Sec.\ \ref{sec:cnt} and \ref{sec:cnv}, for the $C_n+T$ and $\cnv$ insulators respectively. We discuss the utility of our results in interferometric experiments and first-principles calculations in Sec.\ \ref{sec:outlook}.

\section{Review of symmetries in the tight-binding Hamiltonian} \label{app:rvtightbinding}

\begin{figure}[ht]
\centering
\includegraphics[width=7.1 cm]{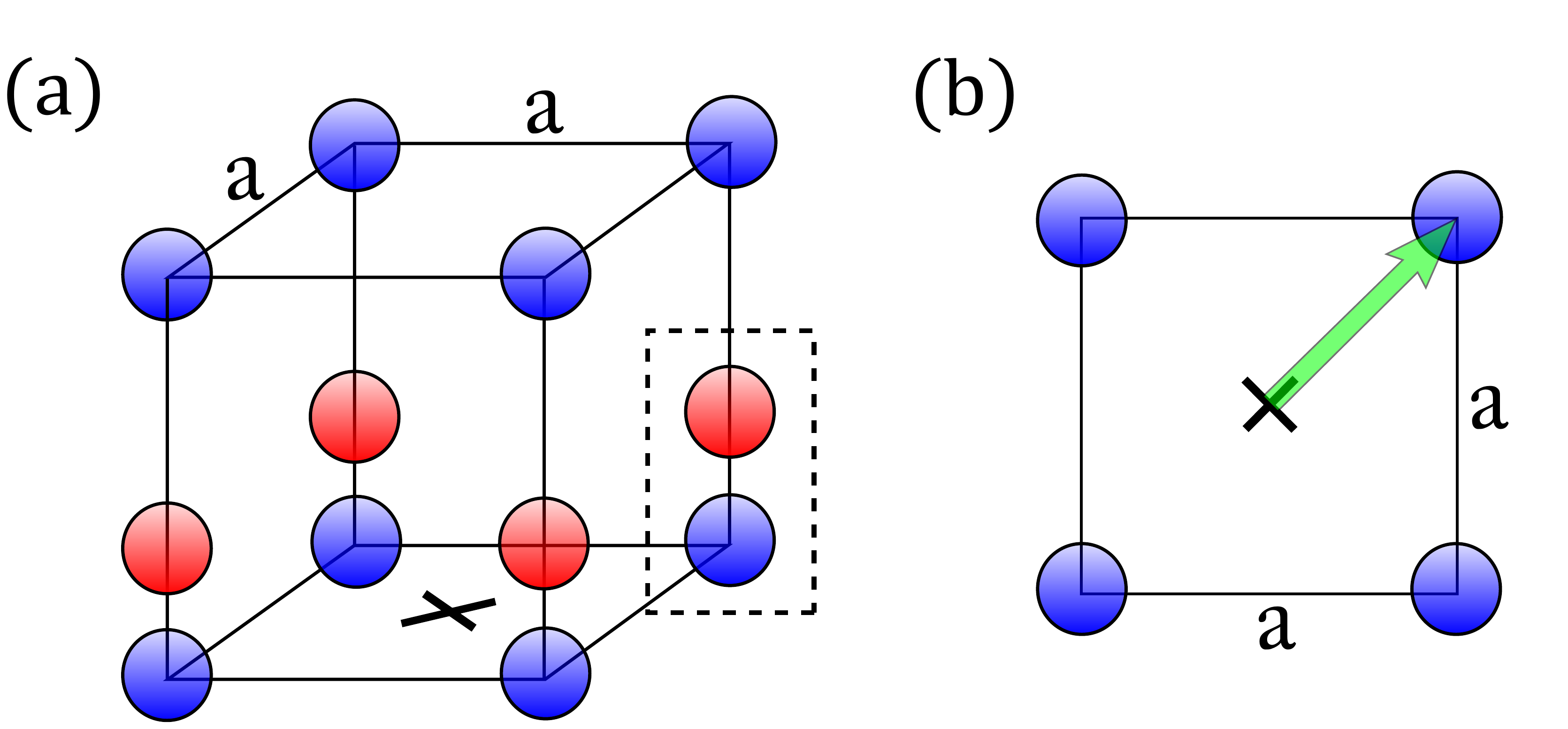}
\caption{ (a) Tetragonal lattice. Each blue sphere corresponds to an atom in one cubic sublattice; red spheres belong to a second cubic sublattice. The unit cell is encircled by a dashed rectangle, and one choice for the spatial origin is the blue atomic site in this rectangle. (b) Top-down view of tetragonal lattice. }\label{fig:tetragonal}
\end{figure}

In the tight-binding method\cite{slater1954,goringe1997,lowdin1950}, the Hilbert space is reduced to a finite number of \low orbitals $\varphi_{\boldsymbol{R},\alpha}$, for each unit cell labelled by the Bravais lattice (BL) vector $\boldsymbol{R}$. For Hamiltonians with discrete translational symmetry, our basis vectors are
\begin{align} \label{basisvec}
\phi_{\boldsymbol{k}, \alpha}(\boldsymbol{r}) = \tfrac{1}{\sqrt{N}} \sum_{\boldsymbol{R}} e^{i\boldsymbol{k} \cdot (\boldsymbol{R}+\boldsymbol{r_{\alpha}})} \pdg{\varphi}_{\boldsymbol{R},\alpha}(\boldsymbol{r}-\boldsymbol{R}-\boldsymbol{r_{\alpha}}),
\end{align}
where $\boldsymbol{k}$ is a crystal momentum, $N$ is the number of unit cells, $\alpha$ labels the \low orbital, and $\boldsymbol{r_{\alpha}}$ denotes the position of the orbital $\alpha$ as measured from the origin in each unit cell. As an example, we consider a tetragonal lattice with a two-atom unit cell -- the lattice is composed of two interpenetrating cubic sublattices, which are correspondingly colored red and blue in Fig.\ \ref{fig:tetragonal}(a). Our tight-binding basis comprises $(p_x,p_y)$ orbitals on each sublattice, which we label by $\alpha =1$ (resp.\ $2$) for $p_x+ ip_y$ (resp. $p_x- ip_y$) orbitals on the blue sublattice, and $ \alpha =3$ (resp.\ $4$) for $p_x- ip_y$ (resp. $p_x+ ip_y$) orbitals on the red sublattice. Suppose the spatial origin lies on an atomic site in the blue sublattice, e.g., the corner of the cube in Fig.\ \ref{fig:tetragonal}(a).  With this choice of origin, the spatial embeddings are $\boldsymbol{r}_1=\boldsymbol{r}_2=0$, and $\boldsymbol{r}_3=\boldsymbol{r}_4 = \Delta \hat{z}$ corresponds to the vertical offset between the two sublattices.\\

The tight-binding Hamiltonian is defined as 
\begin{align} \label{tbHam}
H(\boldsymbol{k})_{\alpha \beta} = \int d^dr\,\phi_{\boldsymbol{k},\alpha}(\boldsymbol{r})^* \,\hat{H} \,\phi_{\boldsymbol{k},\beta}(\boldsymbol{r}), 
\end{align}
where $\hat{H} = p^2/2m + V(\boldsymbol{r})$ is the single-particle Hamiltonian without spin-orbit coupling. The energy eigenstates are labelled by a band index $n$, and defined as $\psi_{n,\boldsymbol{k}}(\boldsymbol{r}) = \sum_{\alpha}  \,u_{n,\boldsymbol{k}}(\alpha)\,\phi_{\boldsymbol{k}, \alpha}(\boldsymbol{r})$, where  
\begin{align} \label{eigentb}
\sum_{\beta} H&(\boldsymbol{k})_{\ab} \,u_{n,\boldsymbol{k}}(\beta)  = \varepsilon_{n,\boldsymbol{k}}\,u_{n,\boldsymbol{k}}(\alpha), \;\; \text{or} \notag \\
&H(\boldsymbol{k})\,\ket{u_{n,\boldsymbol{k}}} = \varepsilon_{n,\boldsymbol{k}}\,\ket{u_{n,\boldsymbol{k}}}.
\end{align} 
Let us define $U_{\sma{2\pi/n}}$ ($R_{\sma{2\pi/n}}$) as the matrix representation of a $C_n$-rotation in the basis of \low orbitals (in $\R^d$, where $d$ is the dimension of the Brillouin zone). Equivalently, the crystal momenta $\boldsymbol{k}$ and $R_{\sma{2\pi/n}}\boldsymbol{k}$ are related by: $R_{\sma{2\pi/n}}(k_x,k_y,k_z) = ( k_x \co 2\pi/n - k_y \si 2\pi/n, k_y \co 2\pi/n + k_x \si 2\pi/n, k_z)$. The Bloch Hamiltonian of a $C_n+T$ insulator has the $C_n$ symmetry\cite{AAchen}:
\begin{align}
 U_{\sma{2\pi/n}}\,H(\,\boldsymbol{k}\,)\,[\,U_{\sma{2\pi/n}}\,]^{\mo} = H(\,R_{\sma{2\pi/n}}\boldsymbol{k}\,),
\end{align}
and the time-reversal symmetry
\begin{align}
T\,H(\,\boldsymbol{k}\,)\,T^{\mo} = H(\,-\boldsymbol{k}\,).
\end{align}
Without spin-orbit coupling, our wavefunctions transform with integral angular momentum. Hence, under $C_n$ rotation, ${[\,U_{\sma{2\pi/n}}\,]}^n=I$; under time-reversal, $T^2 =I$. 

\section{$C_n+T$ insulators} \label{sec:cnt}


\begin{figure}[ht]
\centering
\includegraphics[width=7.1 cm]{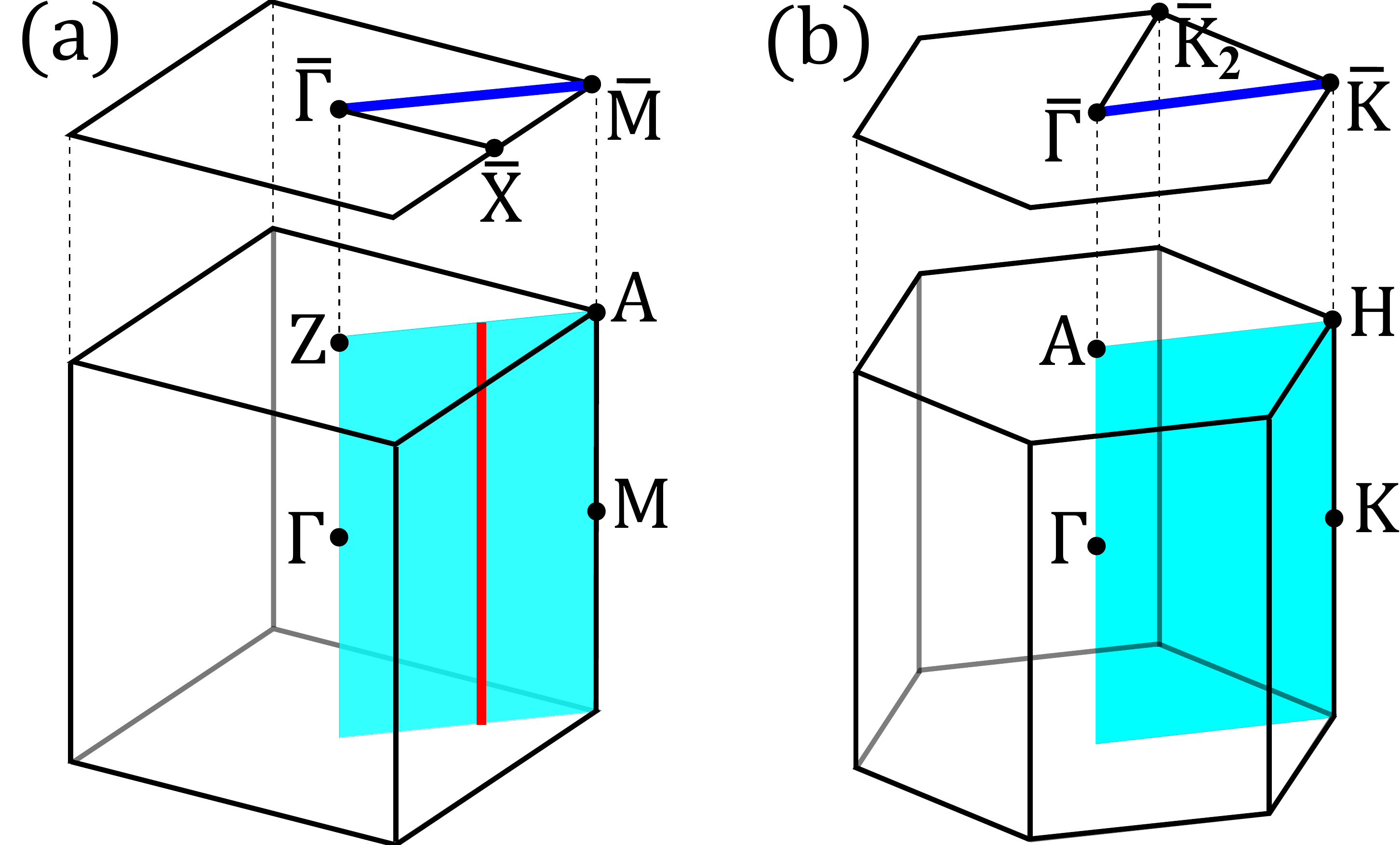}
\caption{ (a)  Bottom: 3D Brillouin zone (BZ) of a simple tetragonal lattice with $C_4$ symmetry. If the tetragonal lattice is symmetric under the larger $C_{4v}$ group, we define the half-mirror plane indicated in blue. Top: 001-surface BZ of the same lattice, with the half-mirror line indicated in blue. (b) 3D BZ (bottom) and 001 BZ (top) of a hexagonal lattice with $C_6$ symmetry. For hexagonal lattices with $C_{6v}$ symmetry, the half-mirror plane and half-mirror line are also indicated.}\label{fig:BZ}
\end{figure}

It is known from Ref.\ \onlinecite{fu2011} that the $C_n+T$ insulator is characterized by two $\Z_2$ indices $\{\Gamma_n(\bar{k}_z)\}$, for $\bar{k}_z \in \{0,\pi\}$; $\Gamma_n(0) \in \{+1,-1\}$ describes the Bloch wavefunctions in the plane $k_z=0$. In analogy with the spin-orbit-coupled $\Z_2$ insulator, $\{\Gamma_n(0),\Gamma_n(\pi)\}$ shall be referred to as weak indices\cite{fu2011,fu2007b,roy2009a}. $\Gamma_n(0)=\Gamma_n(\pi)=1$ ($-1$) corresponds to the trivial (weak) phase, and $\Gamma_n(0)=-\Gamma_n(\pi)$ describes a strong phase. The product $\Gamma_n(0)\,\Gamma_n(\pi)$ is a strong index that determines the absence or presence of robust surface modes on the 001 surface; we take $\hat{z}$ to lie along the principal $C_n$ axis. \\

\begin{figure}[ht]
\centering
\includegraphics[width=8.6 cm]{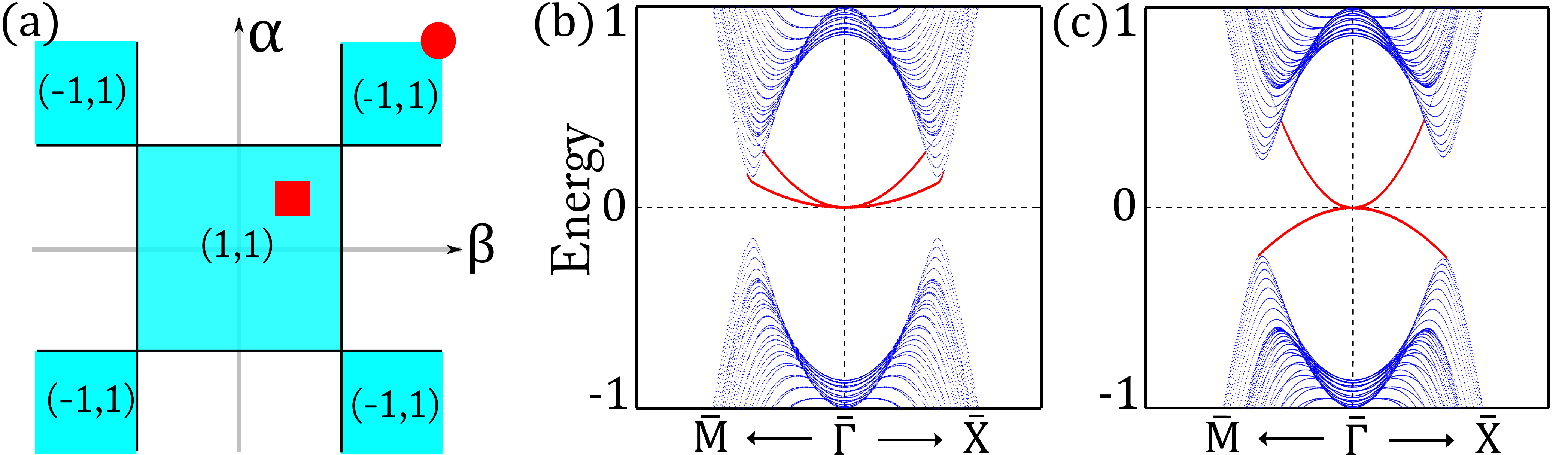}
\caption{(a) Phase diagram of $C_4+T$ model (\ref{cfvmodel}), as parametrized by $\alpha$ and $\beta$; we fix $n_z=1$ and $\delta=0$. Blue (uncolored) regions correspond to gapped (gapless Weyl) phases\cite{AAchen}. The weak indices in each gapped phase are indicated by $(\Gamma_4(0),\Gamma_4(\pi))$. The blue square in the center is approximately bound by $|\alpha|<2$ and $|\beta|<2$. The 001-surface spectrum is plotted for two representative points on the phase diagram: (b) for $\alpha=\beta=1$, and (c) for $\alpha=\beta=4$. $\bar{\Gamma},\bar{M}$ and $\bar{X}$ are high-symmetry momenta defined in Fig.\ \ref{fig:BZ}(a).}\label{fig:c4vmodel}
\end{figure}

We give these weak indices a physically transparent interpretation from the perspective of holonomy. For illustration, we consider a $C_4+T$ model on the tetragonal lattice that is described in Sec.\ \ref{app:rvtightbinding} and illustrated in Fig.\ \ref{fig:tetragonal}(a); ); the corresponding surface and bulk Brillouin zones are illustrated in Fig.\ \ref{fig:BZ}(a). Our tight-binding basis comprises $(p_x,p_y)$ orbitals on each sublattice; each pair transforms in the two-dimensional irreducible representation (irrep) of $C_4+T$. By this, we mean that the $p_x\pm ip_y$ orbitals are individually eigenstates of four-fold rotation, while time-reversal maps $p_x+ip_y \rightarrow p_x-ip_y$. In short, we call such irreps doublets, and we assume that one-dimensional (singlet) irreps are absent in the low-energy, effective Hamiltonian:
\begin{align} \label{cfvmodel}
& H(\boldsymbol{k}) =  \big[\mo + 8\,f_1(\boldsymbol{k})\,\big]\,\Gamma_{03}  + \big[\,2\,f_2(\boldsymbol{k})+\delta\,f_3(\boldsymbol{k})\,\big]\,\gam{11} \notag \\
&\;\;\;\;  + \alpha\,f_4(\boldsymbol{k})\,\gam{01}  +  \beta\,f_5(\boldsymbol{k})\,\gam{32} + 2\,f_6(\boldsymbol{k})\,\gam{12},
\end{align}
where $f_1 = 3-\text{cos}(k_x)-\text{cos}(k_y)-\text{cos}(n_zk_z)$, $f_2 = 2-\text{cos}(k_x)-\text{cos}(k_y)$, $f_3=\text{cos}(k_z)$, $f_4 = \text{cos}(k_y)-\text{cos}(k_x)$,  $f_5 = \text{sin}(k_x)\,\text{sin}(k_y)$ and $f_6 = \text{sin}(n_zk_z)$. In $\Gamma_{ab} = \sigma_a \otimes\tau_b$, $\sigma_i$ and $\tau_i$ are Pauli matrices for $i \in \{1,2,3\}$, while $\sigma_0$ and $\tau_0$ are identities in each 2D subspace.  $\ket{\sigma_3= \pm 1,\tau_3=+1}$  label $\{p_x \pm i p_y\}$ orbitals on one sublattice, and $\ket{\sigma_3= \pm 1,\tau_3=-1}$  label $\{p_x \mp i p_y\}$ orbitals on the other. This Hamiltonian is four-fold symmetric: $\Gamma_{33}\,H(k_x,k_y,k_z)\,\Gamma_{33} = H(\,-k_y,k_x,k_z\,)$, and time-reversal symmetric: $\Gamma_{10}\,H(\boldsymbol{k})^*\,\Gamma_{10} = H(-\boldsymbol{k})$. The ground state of (\ref{cfvmodel}) comprises its two lowest-lying bands. The phase diagram of this model is plotted in Fig.\ \ref{fig:c4vmodel}(a) for different parametrizations of (\ref{cfvmodel}). \\

To probe the bulk topology, we perform parallel transport along a bent loop that connects two $C_4$-invariant points; $C_n$-invariant points refer to momenta which are invariant under an $n$-fold rotation, up to a reciprocal lattice vector. We define $l_4(0)$ as the loop connecting $M-\Gamma-M$ in the $k_z=0$ plane, and $l_4(\pi)$ connects $A-Z-A$ in the $k_z=\pi$ plane. They are respectively depicted by red and brown lines in Fig.\ \ref{fig:BZsWloops}(a).  As we review in App.\ \ref{app:rvWilson}, the matrix representation of holonomy is known as the Wilson loop, and it is the path-ordered exponential of the Berry-Wilczek-Zee connection\cite{wilczek1984,berry1984} $A(\boldsymbol{k})_{ij} = \bra{u_{i,\boldsymbol{k}}}\,\nabla_{\boldsymbol{k}}\,\ket{u_{j,\boldsymbol{k}}}$: 
\begin{align} \label{wloopdifferentiable}
\W[l] = \text{exp}\,\big[{-\int_l} dl \cdot A(\boldsymbol{k})\,\big].
\end{align}
Here, $\ket{u_{j,\boldsymbol{k}}}$ is an occupied eigenstate of the Bloch Hamiltonian, as defined in\ (\ref{eigentb}); $l$ denotes a loop and $A$ is a matrix with dimension equal to the number ($\noc$) of occupied bands. The gauge-invariant spectrum of $\W[l]$ is also known as the Berry-Zak phase factors ($\{\text{exp}(i \vartheta)\}$)\cite{zak1982,*zak1989,AA2014}.  Let us show that the spectrum of $\W[l_4(\bar{k}_z)]$ encodes the weak index $\Gamma_4(\bar{k}_z)$. If we define $d_{4}$ is the number of $-1$ eigenvalues in the spectrum of $\W[l_{4}]$, then the weak indices $\{\Gamma_4(0),\Gamma_4(\pi)\}$ are related to $\{d_{4}(0),d_4(\pi)\}$ by
\begin{align} \label{2dz2inv}
\Gamma_n(\bar{k}_z) = i^{d_n(\bar{k}_z)} \in \{1,\mo\}; \;\; \;\bar{k}_z \in \{0,\pi\}
\end{align}
for $n=4$; as we will shortly clarify, $d_4$ is necessarily even. This weak index is equivalent to an alternative formulation in Ref.\ \onlinecite{fu2011}, where it is expressed as an invariant involving the Pfaffian of a matrix; cf. App.\ \ref{app:equivalencepfaffian}. \\

We provide a geometrical interpretation of\ (\ref{2dz2inv}): the parity of $d_4(\bar{k}_z)/2$ specifies one of two classes of a special rotation, which is in one-to-one correspondence with two sectors of ground states in the $k_z=\bar{k}_z$ plane. The following discussion briefly clarifies the nature of this rotation; further details may be found in App.\ \ref{app:specifickz}. Due to two-fold rotational and TRS, we can choose a basis in which $\W[l_4(\bar{k}_z)] \in SO(\noc)$, i.e., they are proper rotations in $\R^{\noc}$. Since the bands derive from doublet orbitals, $\noc$ is even. A rotation $R$ in $\noc=2m$ dimensions is described by $m$ invariant planes, and an angle of rotation in each plane. If all $m$ angles equal to $\theta$, such a rotation is called equiangular -- there is an invariant plane through any arbitrary vector of space and all vectors are rotated by the same angle $\theta$\cite{Gelfand}. In an appropriate basis, $\W[l_4(\bar{k}_z)]$ is a product of two equiangular rotations, each of angle $\pi/2$ -- the net effect is that a vector may be maximally rotated by angle $\pi$. The set of vectors which are rotated by $\pi$ is defined as the maximally-rotated subspace, and we interpret $d_{4}(\bar{k}_z)$ in (\ref{2dz2inv}) as the dimension of this subspace. These vectors always come in pairs, since each eigenvalue of an even-dimensional rotation has a complex-conjugate partner\cite{Gelfand}. There are then two classes of $\W$ distinguished by the parity of $d_{4}(\bar{k}_z)/2$; in the nontrivial (trivial) class an odd (even) number of pairs are maximally rotated. The simplest example for $\noc=2$ is the equiangular rotation $R_{\pm} = e^{\pm i \sigma_2\pi/2}$. $\W$ is either the trivial identity: $R_+R_-=I$, or it rotates any vector by $\pi$: $R_{\pm}R_{\pm}=-I$. \\

Now we demonstrate how to realize both classes of $\W$ in the model of (\ref{cfvmodel}). Let us consider a family of loops $\{l_4(k_z)\}$ in planes of constant $k_z$, such that $l_4(0)$ is the red line in Fig.\ \ref{fig:BZsWloops}(a), and all other loops project to $l_4(0)$ in $\hat{z}$. In the trivial phase (parametrized by $\alpha=\beta=1$), we find $\W[l_4(0)]=\W[l_4(\pi)]=I$, or equivalently $\Gamma_4(0)=\Gamma_4(\pi)=+1$. The eigenvalues of $\W[l_4(k_z)]$ interpolate between $\{1,1\}$ (at $k_z=0$) to $\{1,1\}$ (at $k_z=\pi$), as illustrated in Fig.\ \ref{fig:BZsWloops}(b). The absence of surface modes on the line $\bar{M}-\bar{\Gamma}-\bar{X}$ is demonstrated in Fig.\ \ref{fig:c4vmodel}(b). In comparison, the strong phase ($\alpha=\beta=4$) is characterized by $\W[l_4(0)]=-\W[l_4(\pi)]=-I$, or $\Gamma_4(0)=-\Gamma_4(\pi)=-1$; its surface modes are illustrated in Fig.\ \ref{fig:c4vmodel}(c). As $k_z$ is varied from $0$ to $\pi$ in Fig.\ \ref{fig:BZsWloops}(c), the Berry phases $\{\vartheta(k_z)\}$ interpolate across the maximal range $(-\pi,\pi]$ -- we call this property spectral flow. In comparison, the weak phase is identified by $\W[l_4(0)]=\W[l_4(\pi)]=-I$, which clearly does not exhibit spectral flow.\\

\begin{figure}[ht]
\centering
\includegraphics[width=8.6 cm]{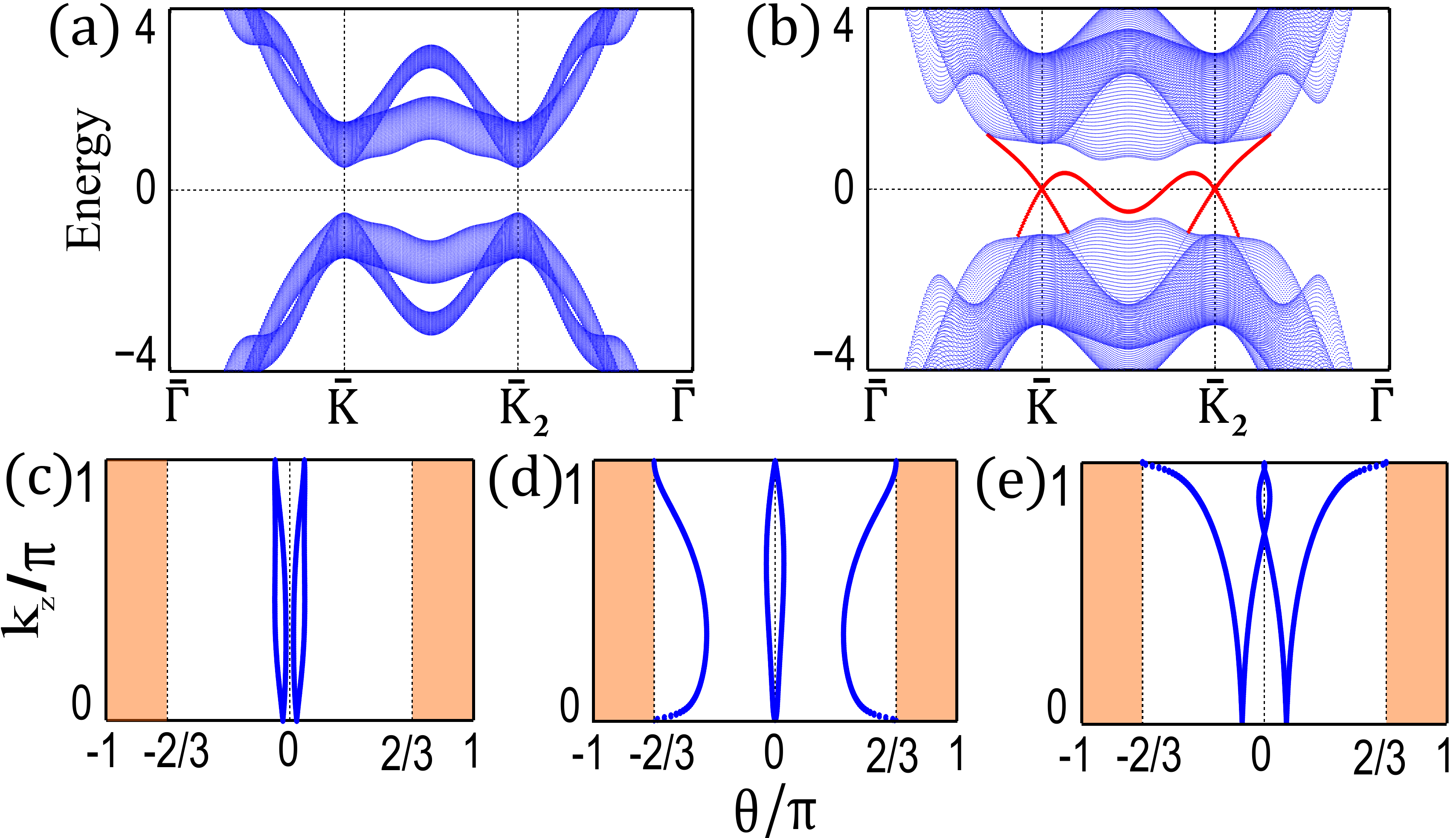}
\caption{ Characterization of the $C_6+T$ insulator.(a) (resp.\ (b)) is the $001$-surface spectrum of a trivial (resp.\ strong) insulator. $\bar{\Gamma}$,$\bar{K}$ and $\bar{K}_2$ are $C_3$-invariant momenta defined in Fig.\ \ref{fig:BZ}(b). (c), (d) and (e) respectively illustrate the Berry phases of  trivial, weak and topological phases. }\label{fig:C6Tcomposite}
\end{figure}

The story of the $C_6+T$ insulator proceeds analogously. We provide a model whose details are reported in App.\ \ref{app:modelC6T}; the trivial and strong phases of this model are distinguished by surface modes, as illustrated in Fig.\ \ref{fig:C6Tcomposite}(a) and (b). The bulk topology is unveiled by the following bent loops, which are illustrated in Fig.\ \ref{fig:BZsWloops}(d): we define $l_6(0)$ as the loop  $K-\Gamma-K$ (red), and $l_6(k_z)$ as the $\hat{z}$-projection of $l_6(0)$ in the plane of constant $k_z$\cite{footnote1}. For $\bar{k}_z \in \{0,\pi\}$, a basis may be found where $\W[l_6(\bar{k}_z)]$ is a product of two equiangular rotations, each of angle $\pi/3$ -- a vector may be maximally rotated by an angle $2\pi/3$, as we show in App.\ \ref{app:specifickz}. (\ref{2dz2inv}) similarly applies for $n=6$, if we define $d_6$ as the dimension of the maximally-rotated subspace. Equivalently, $d_{6}/2$ is the number of exp$(i2\pi/3)$-eigenvalues in the spectrum of $\W[l_6(\bar{k}_z)]$. Like $\Gamma_4$, $\Gamma_6$ is also equivalent to a Pfaffian invariant, though the proposed formula in Ref.\ \onlinecite{fu2011} requires a clarification; cf. App.\ \ref{app:equivalencepfaffian}. In the strong phase, the Berry phases interpolate across the maximal range $[-2\pi/3,2\pi/3]$; compare Fig.\ \ref{fig:BZsWloops}(e) with Fig.\ \ref{fig:BZsWloops}(f). \\

Beyond these two models, we would like to generalize our results to insulators with any number of occupied doublet bands. For $n \in \{4,6\}$, the spectrum of $\W[l_n(\bar{k}_z)]$ falls into two classes which are labelled by the weak index $\Gamma_n(\bar{k}_z) \in \{\pm 1\}$; the structure of the two classes is laid out in Tab.\ \ref{C4spectrumtable}. This $\Z_2$ classification of $\W[l_n(\bar{k}_z)]$ relies essentially that all $\noc$ bands transform in the doublet irrep of $C_n+T$.  If singlet bands were included in the calculation of $\W$, we note that the $\W$-eigenvalues are not robustly quantized, e.g., for $\Gamma_n=-1$ (second and fourth rows in Tab.\ \ref{C4spectrumtable}), the $\W$-spectrum will deviate from exp$(\pm i 4\pi/n)$ due to hybridization between singlet and doublet bands. Our conclusion that only doublet bands admit a $\Z_2$ classification is consistent with Ref.\ \onlinecite{fu2011}.\\

\begin{table}[H]
	\centering
		\begin{tabular} {|c|c|c|} \hline
		  $\noc$ & $\Gamma_n$ & Spectrum of $\W[l_n(\bar{k}_z)]$ \\ \hline
			$4m$  & $1$ & $\{\lambda_1\}_4, \{\lambda_2\}_4,\;\ldots\;,\{\lambda_m\}_4$  \\ \cline{2-3}
			& $\mo$ & $e^{i4\pi/n},\,e^{-i4\pi/n},\,1,\,1,\, \{\lambda_1\}_4, \;\ldots\;,\{\lambda_{m-1}\}_4$ \\ \hline
			$4m+2$  & $1$ & $ 1,\,1,\,\{\lambda_1\}_4, \;\ldots\;,\{\lambda_m\}_4 $ \\ \cline{2-3}
			& $\mo$ & $e^{i4\pi/n},\,e^{-i4\pi/n},\,\{\lambda_1\}_4, \;\ldots\;,\{\lambda_m\}_4$ \\ \hline
			\end{tabular}
		\caption{ Spectrum of the bent Wilson loop. We consider two cases: (i) the number of occupied bands ($\noc$) is $4m$, for non-negative integer $m$, and (ii) $\noc=4m+2$. In either case, the spectrum has two possible structures, as labelled by $\Gamma_n \in \{+1,-1\}$. $\{\lambda_1\}_4$ denotes a doubly-degenerate eigenvalue and its complex conjugate: $\{\lambda_1,\lambda_1,\lambda_1^*,\lambda_1^*\}$. This table is derived in App.\ \ref{app:specifickz}. \label{C4spectrumtable}}
\end{table}

In the strong phase, $\Gamma_n(0)=-\Gamma_n(\pi)$ is a sufficient condition for spectral flow: the Berry phases $\{\vartheta(k_z)\}$ robustly interpolate across the full range $[-4\pi/n,4\pi/n]$, in the interval $k_z \in [0,\pi]$. The converse is also true for the trivial and weak phases: $\Gamma_n(0)=\Gamma_n(\pi)$ implies the lack of spectral flow. Though a proof can be written, we prefer to make a pictorial argument through Fig.\ \ref{fig:C6Tcomposite}(c) to (e), where we compare trivial, weak and strong phases in a $C_6+T$ model with four occupied bands. Crucial to this argument is that $\{\vartheta(k_z)\}$ satisfy certain symmetry constraints: (i) for any $k_z \in [0,\pi]$, the spectrum of $\W$ only comprises complex-conjugate pairs $\{\text{exp}(i \vartheta),\text{exp}(\text{-}i \vartheta)\}$, as proven in App.\ \ref{app:generalkz}, and  (ii) all $\W$-eigenvalues are doubly-degenerate at $k_z=0$ and $\pi$ (cf. Tab. \ref{C4spectrumtable}). \\

We remark on the well-known $U(1)$ ambiguity of the Wilson loop originating from the choice of real-spatial origin.\cite{AA2014} In our $C_6+T$ study, there is a unique spatial origin, modulo Bravais-lattice translations, which is invariant under six-fold rotation; equivalently, there exists only a single Wyckoff position with unit multiplicity, and we assume this point as our spatial origin throughout this paper. In our $C_4+T$ study, there are instead two possible choices for the spatial origin which are each invariant under four-fold rotation: either the corner or center of the square unit cell in Fig.\ \ref{fig:tetragonal}(b). Translating the spatial origin from corner to center induces a global phase shift of all $\W$-eigenvalues by $\bG {\cdot} \delta \br$,\cite{AA2014} with $\bG$ the reciprocal vector connecting base and end points of the quasimomentum loop, and $\delta \br$ the real-spatial vector connecting corner to center [as illustrated by the green arrow in Fig.\ \ref{fig:tetragonal}(b)]; as shown in  App.\ \ref{app:origin}, $\bG \cdot \delta \br=\pi$, and consequently the weak indices are modified as $\Gamma_4(0) \rightarrow -\Gamma_4(0)$ and $\Gamma_4(\pi)\rightarrow -\Gamma_4(\pi)$, though their product $\Gamma_4(0)\Gamma_4(\pi)$, the strong index, is invariant. To distinguish between two topologically-distinct bandstructures in the $k_z=0$ plane, one therefore has to compute $\Gamma_4(0)$ using the same choice of origin. We further remark on the experimental measurability of our weak indices in Sec.\ \ref{sec:outlook}.\\

Finally, we point out an alternative characterization of the $C_n+T$ insulator by Berry phases, which was described in Ref. \onlinecite{Maryam2014} for a different choice of loop. Their characterization is useful to identify the strong index: the product $\Gamma_n(0)\Gamma_n(\pi)$, but cannot individually distinguish the weak indices: $\Gamma_n(0)$ and $\Gamma_n(\pi)$. 

\section{$C_{nv}$ insulators} \label{sec:cnv}

\begin{figure}[ht]
\centering
\includegraphics[width=8.6cm]{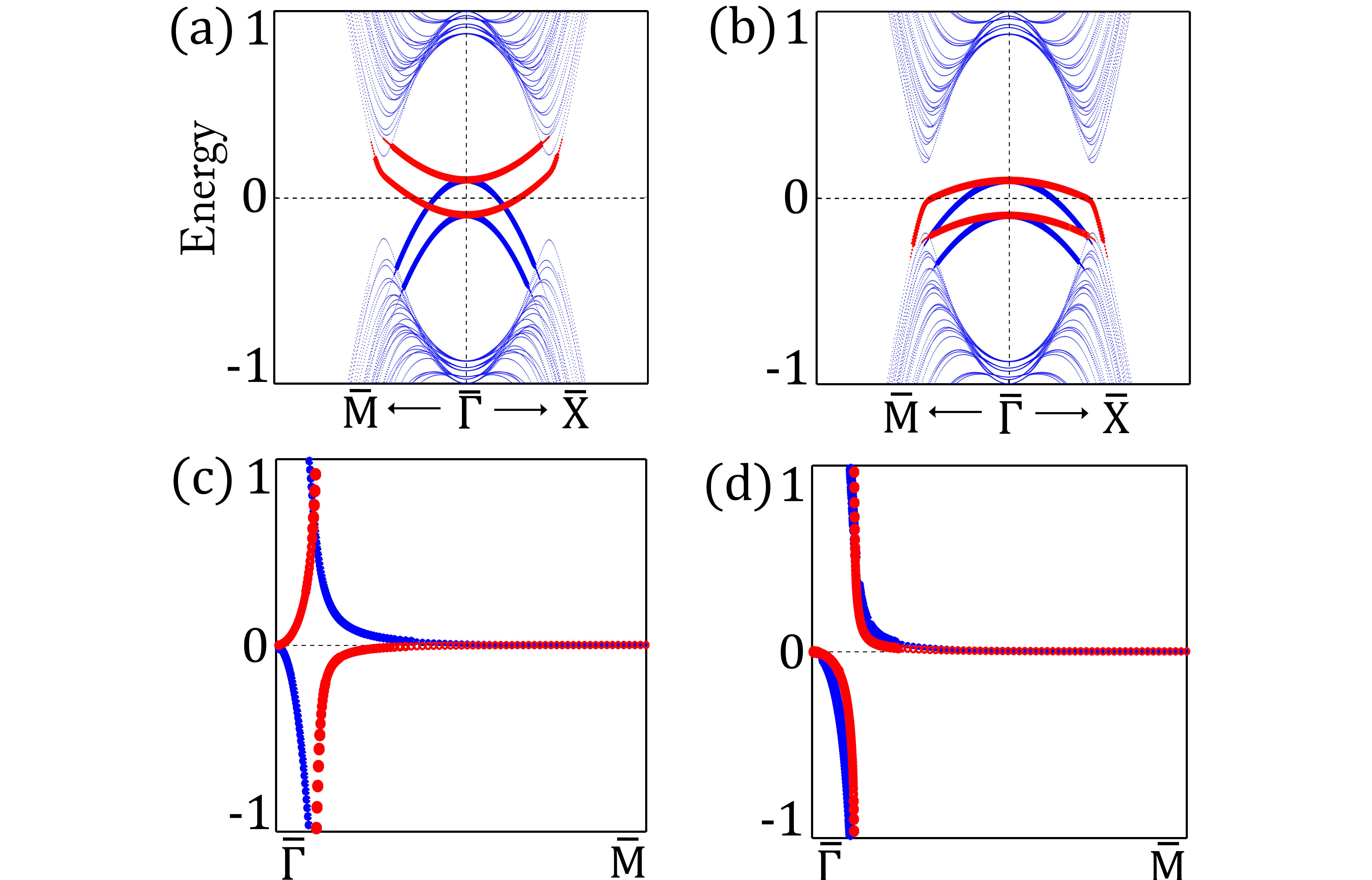}
\caption{ (a) (resp.\ (c)) is the 001-surface dispersion (resp.\ Berry-phase spectrum) of the $C_{4v}$ topological insulator, with $\chi=2$. (b) and (d) describe a trivial $\cfv$ insulator with $\chi=0$. Along the HML connecting $\bar{\Gamma}$ and $\bar{M}$, mirror-even (resp.\ odd) surface bands are highlighted in red (resp.\ blue). Similarly, the Berry phase of the mirror-even (resp.\ odd) subspace is colored red (resp.\ blue).}\label{fig:CNVcomposite}
\end{figure}

In our second case study, we aim to express the halved-mirror chirality, an integer invariant that characterizes $\cnv$ insulators, in terms of Berry phases. For illustration, we employ the $\cfv$-symmetric model of Eq. (\ref{cfvmodel}); in addition to the above-mentioned symmetries, the model is also symmetric under reflection: $\Gamma_{23} \,H(k_x,k_y,k_z)\,\gam{23} = H(k_y,k_x,k_z)$. While the halved chirality is well-defined and robust without TRS, we nevertheless keep TRS for simplicity. The parameters $n_z=2$, $\delta=0.1$ and $\alpha=\beta=4$ correspond to the 001-surface dispersion in Fig.\ \ref{fig:CNVcomposite}(a). Let us define the half-mirror line (HML)\cite{AAchen} as connecting $\bar{\Gamma}$ and $\bar{M}$ in the 001-surface BZ, as drawn in Fig.\ \ref{fig:BZ}(a). Each state in the HML may be labelled by its eigenvalue under the reflection $(x,y)\rightarrow (y,x)$, which is represented by the operator $-\gam{23}$; in short, we call bands with eigenvalue $+1$ $(-1)$ as mirror-even (mirror-odd). The surface bands are characterized by  the {halved}-mirror chirality $\pdg{\chi} \in \Z$, which we define as the {difference} in number of mirror-even chiral modes  with mirror-odd chiral modes within the HML. To extract $\chi$ from the surface dispersion, draw a constant-energy line that is within the energy gap and parallel to the HML, e.g., we pick the zero-energy line in Fig.\ \ref{fig:CNVcomposite}(a). Let us parametrize the HML by $\kpar \in s\pi(\hat{x}+\hat{y})$, where $s=0$ $(1)$ at $\bar{\Gamma}$ ($\bar{M}$). At each intersection with a surface band, calculate [sign of the group velocity $dE/ds$] $\times$ [mirror eigenvalue]; sum this quantity over all intersections along the HML to obtain $\pdg{\chi}$. In our example, the two intersections result in $\pdg{\chi} = (1)(1) + (-1)(-1)=2$. \\

Now we describe how the surface-band index ${\chi}$ is encoded in the {bulk} wavefunctions. Taking $\hat{z}$ to lie along the rotational axis, the HML in the surface BZ is the $\hat{z}$-projection of a half-mirror-plane (HMP) in the 3D BZ, as illustrated in Fig.\ \ref{fig:BZ}(a). Let us parametrize HMP by $t \in [0,1]$ and $k_z \in (-\pi,\pi]$; $t=0$ $(1)$ along the first (second) $C_4$-invariant line. Then the halved chirality has the following expression by bulk wavefunctions:
\begin{align} \label{mirrorchirality}
\pdg{\chi} = \frac{1}{2\pi}\,\int_{\text{HMP}} dt\,dk_z\,(\,\calf_{e} - \calf_{o}\,).
\end{align}
$\calf_{e}$ ($\calf_{o}$) is defined as the Berry curvature of occupied doublet bands\cite{berry1984,zak1982,zak1989}, as contributed by the mirror-even (-odd) subspace. To express (\ref{mirrorchirality}) in terms of Berry phases, we consider a \emph{different} family of non-contractible loops $\{z(\kpar)\}$ which lie within the HMP; an example of a loop is illustrated in red in Fig.\ \ref{fig:BZ}(a). To compare, the previous loop $l_n(k_z)$ lies in a plane of constant $k_z$, while $z(\kpar)$ lies on a line of fixed $\kpar \in$ HML. Denoting the eigenvalues of $\W[z(\kpar)]$ by $\{\text{exp}(i\vartheta(\kpar))\}$, we plot the Berry phases $\{\vartheta\}$ as a function of $\kpar$ in Fig.\ \ref{fig:CNVcomposite}(c). Due to the orthogonality of the mirror subspaces, we may label each branch of $\vartheta$ by its mirror eigenvalue: $\vartheta_e$ ($\vartheta_o$) in the even (odd) subspace is colored red (blue). By Stoke's theorem, we rewrite (\ref{mirrorchirality}) as the net change in $\vartheta_e$ in the interval $s \in [0,1]$, minus the net change in $\vartheta_o$:
\begin{align} \label{winding}
\pdg{\chi} = \frac{1}{2\pi}\,\int_0^1 ds\,\bigg(\,\partdif{\vartheta_e}{s} - \partdif{\vartheta_o}{s}\,\bigg).
\end{align}
Since $\vartheta_e=\vartheta_o$ at $s=0$ and $s=1$\cite{AAchen}, $\chi$ is quantized to integers -- each nonzero integer characterizes a topologically distinct type of spectral flow. $\chi$ may be extracted from $\{\vartheta\}$ in a manner that is analogous to the surface-band index: by considering the intersections of $\{\vartheta\}$ with an arbitrary constant-phase line. At each intersection, we evaluate [sign of $d\vartheta/ds$] $\times$ [mirror eigenvalue], then sum this quantity over all intersections along the HML. By inspection of Fig.\ \ref{fig:CNVcomposite}(c), we find $\chi=2$, in one-to-one correspondence with the surface-band formulation of $\chi$. For comparison, we plot the surface bands and Berry phases of a trivial insulator in Fig.\ \ref{fig:CNVcomposite}(b) and (d); these are obtained from\ (\ref{cfvmodel}) with parameters $n_z=2$, $\delta=0.1$ and $\alpha=\beta=1$. The formula (\ref{winding}) is applicable to the case of two occupied bands, as in the model\ (\ref{cfvmodel}); to generalize to $2m$ occupied bands for $m>1$, we interpret $\vartheta_e$ ($\vartheta_o$) as the sum of all Berry phases in the even (odd) subspace.  While we have focused on one HMP for the purpose of illustration, the $\cfv$ insulator is characterized by another halved chirality, which is defined on a different HMP. A full discussion of the various HMP's for all relevant $\cnv$ groups is provided in Ref.\ \onlinecite{AAchen}.


\section{Discussion and experimental outlook} \label{sec:outlook}

Wilson loops (synonymously, Berry-Zak holonomies) are widely applied by first-principles calculators to identify topological matter of both gapped\cite{yu2011,alexey2011,Maryam2014} and gapless\cite{tilted_dirac} varieties. Underlying the Wilson-loop method is a group-cohomological classification of quasimomentum manifolds\cite{Cohomological} -- this provides a unifying framework to classify chiral topological insulators,\cite{Haldane1988} and all topological insulators with robust edge states protected by space-time symmetries. Here, we refer to topological insulators with either symmorphic\cite{Classification_Chiu,fu2011,AAchen} or nonsymmorphic spatial symmetries\cite{ChaoxingNonsymm,unpinned,Shiozaki2015,Hourglass,Nonsymm_Shiozaki,mobius_kondo}, the time-reversal-invariant quantum spin Hall phase,\cite{kane2005B} and magnetic topological insulators.\cite{moore2010,fang2013,liu2013,magnetic_ti} The Wilson loop has also proven useful in classifying topological insulators without edge states.\cite{AA2014,hughes2011,turner2012} \\

All of these previous works characterize band insulators by Berry-Zak holonomies along straight loops in momentum space; the Berry-Zak phases are then related to the spatial positions of (hybrid) Wannier functions\cite{AA2014,soluyanov2011}; this relation underlies a bulk-boundary correspondence\cite{Cohomological,fidkowski2011,ZhoushenHofstadter,Maryam2014} between the Berry-Zak phases (a bulk characterization) and robust edge states. Part of this work demonstrates how straight Wilson loops characterize topological phases with $\cnv$ symmetry, and also describes the bulk-boundary correspondence for this symmetry class. \\

Our work also points out that some Berry-Zak phases have no interpretation in terms of polarization, but they are nevertheless interesting characterizations of band insulators. We have shown that bent Wilson loops topologically distinguish between classes of time-reversal-invariant bands with the $C_n$ point-group symmetry; some of these classes do not have robust surface modes, but are experimentally distinguishable by their Berry-Zak phases.\\

The Berry-Zak phase is now measurable by Ramsey interference in cold-atomic systems, for both abelian\cite{atala2013} and non-abelian\cite{Tracy} Wilson loops. In the latter experiment,\cite{Tracy} we emphasize that it is the phase \emph{difference} in the two eigenvalues of the $U(2)$ Wilson loop which is measured, rather than the absolute phase of each eigenvalue. Such an experiment would ideally probe the $C_6+T$ weak index that we propose for two-band systems: the two topologically-distinct classes of bandstructures are distinguished by a phase difference of $0$ and $2\pi/3$ (mod $2\pi$).\\

On the other hand, the $C_4+T$ weak index cannot immediately be measured with the same techniques in Ref.\ \onlinecite{Tracy}, since, in both topologically-distinct classes, the phase difference between the two Wilson-loop eigenvalues vanishes (mod $2\pi$). It is conceivable that a topological phase transition between these two classes is measurable, in analogy with the experiment of Ref.\ \onlinecite{atala2013}. To elaborate on this experiment, Atala \emph{et}.\ \emph{al}.\ measured the quantized \emph{change} in Berry-Zak phase, corresponding to a topological phase transition between two classes of one-dimensional, centrosymmetric bands.\cite{zak1989}\\

\begin{center}
\textbf{{Acknowledgements}}
\end{center}
Chen Fang and Matthew Gilbert played an essential role in developing the theory of $C_{nv}$ insulators. We are grateful to Xi Dai, and the hospitality of IOP China. AA and BAB were supported by NSF CAREER DMR-095242, ONR-N00014-11-1-0635, ARO MURI on topological insulators, grant W911NF-12-1-0461, NSF-MRSEC DMR-1420541, Packard Foundation, Keck grant, “ONR Majorana Fermions” 25812-G0001-10006242-101, and Schmidt fund 23800-E2359-FB625. \\

\begin{center}
\textbf{
{APPENDIX}}
\end{center}

\vspace*{3mm}

Organization of the appendix:  in App.\ \ref{app:modelC6T}, we provide the details of a $C_6+T$ model that has been employed in Sec.\ \ref{sec:cnt}. In App.\ \ref{app:rvWilson}, we review general properties of the Wilson loop. In App.\ \ref{app:generalkz}, we derive certain properties of the bent Wilson loop $\W[l_n(k_z)]$, as applied to the $C_n+T$ insulator; these properties apply for any $k_z$ in the Brillouin zone.  In App.\ \ref{app:specifickz}, we focus on $\W[l_n(\bar{k}_z)]$ in the high-symmetry planes defined by $\bar{k}_z=0$ and $\pi$. Specifically, we develop a geometric interpretation of $\W[l_n(\bar{k}_z)]$ as a special type of proper rotation, and derive the structure of its spectrum. In App.\ \ref{app:equivalencepfaffian}, we show an alternative formulation of the topological invariant $\Gamma_n$, to make contact with previous work in Ref.\ \onlinecite{fu2011}. Finally in App.\ \ref{app:origin}, we describe how the choice of the spatial origin affects the Berry-Zak phases, and describe a translational-invariant formulation of the $C_n+T$ topological invariants.
\appendix 

\begin{widetext}
\section{Model of $C_6+T$ insulator} \label{app:modelC6T}

We model a $C_6+T$ insulator on a hexagonal Bravais lattice; this model is a generalization of the $C_4+T$ model proposed in Ref. \onlinecite{fu2011}. The unit cell comprises two inequivalent atoms $A$ and $B$ along the $c$-axis, and each atom belongs to a triangular lattice -- the entire crystal may be thought of as stacked bilayers of triangular lattices. The orbitals on each atom transform in the $\{p_x,p_y\}$ representation. Our basis is spanned by the Pauli matrices $\tau_i$ and $\sigma_i$: the A-sublattice (B-sublattice) corresponds to $\tau_3=+1$ $(-1)$, and the $p_x$ ($p_y$) orbital corresponds to $\sigma_3=+1$ ($-1$).
  The Hamiltonian may be written as
\begin{align} \label{c6pTmodel}
H(\boldsymbol{k}) =  \begin{pmatrix} h_{AA}(\boldsymbol{k}) && h_{AB}(\boldsymbol{k}) \\ \\ \dg{h_{AB}(\boldsymbol{k})} && h_{BB}(\boldsymbol{k}) \end{pmatrix},
\end{align}
such that $h_{aa}$ acts within the sublattice $a \in \{A,B\}$. We consider sigma-type hopping within each sublattice: nearest-neighbor (next-nearest-neighbor) hoppings are parametrized by $t_1^a$ ($t_2^a$).
\begin{align}
h_{aa}(\boldsymbol{k}) =  & t_1^a\;\begin{pmatrix} 2 \co k_1 + \half \co k_2 + \half \co (k_1-k_2) && \tfrac{\sqrt{3}}{2} \co k_2 - \tfrac{\sqrt{3}}{2} \co (k_1-k_2) \\ \\ \tfrac{\sqrt{3}}{2} \co k_2 - \tfrac{\sqrt{3}}{2} \co (k_1-k_2)  && \tfrac{3}{2} \co k_2 + \tfrac{3}{2} \co (k_1-k_2)
 \end{pmatrix}  \notag \\
 & + t_2^a\; \begin{pmatrix} \tfrac{3}{2} \co (k_1+k_2)+ \tfrac{3}{2} \co (2 k_1-k_2) && \tfrac{\sqrt{3}}{2} \co (k_1+k_2) - \tfrac{\sqrt{3}}{2} \co (2 k_1-k_2) \\ \\ \tfrac{\sqrt{3}}{2} \co (k_1+k_2) - \tfrac{\sqrt{3}}{2} \co (2 k_1-k_2) && 2 \co (2 k_2 - k_1) + \half \co (k_1+k_2) + \half \co (2 k_1-k_2)
 \end{pmatrix}.
\end{align}
Here, we have chosen the non-orthogonal coordinates $\{k_1,k_2,k_z\}$, defined by $\boldsymbol{k} = (\,k_1 \boldsymbol{b_1} + k_2 \boldsymbol{b_2} + k_z \boldsymbol{b_3}\,)/2\pi$; $k_1,k_2,k_z \in [0,2\pi)$; the reciprocal lattice vectors are $\boldsymbol{b_1} = b/2\,(\sqrt{3},\mo,0)$, $\boldsymbol{b_2} = b\;(0,1,0)$ and $\boldsymbol{b_3}= 2\pi/c\,(0,0,1)$ for some parameters $b$ and $c$. \\

We include inter-sublattice orbital-independent hoppings: nearest-neighbor (next-nearest-neighbor) hoppings within a bilayer are parametrized by $t_1'$ ($t_2'$); nearest-neighbor hoppings between bilayers are parametrized by $t_z'$.
\begin{align}
h_{AB}(\boldsymbol{k}) = \big(\;t_1' + 2 t_2'\,\big(\co k_1 + \co k_2 + \co (k_1-k_2)\big) + t_z'\,e^{ i k_z} \;\big)\;I.
\end{align}
This Hamiltonian has the time-reversal symmetry: $H(\boldsymbol{k})^* = H(-\boldsymbol{k})$, and six-fold symmetry: $U_{\sma{\pi/3}}\,H(\boldsymbol{k})\,U_{\sma{\pi/3}}^{\mo} = H(\,R_{\sma{\pi/3}}\boldsymbol{k}\,)$, for $U_{\sma{\pi/3}} = \text{exp}[-i \sigma_2 \pi/3]$; $\boldsymbol{k}$ and $R_{\sma{\pi/3}}\boldsymbol{k}$ are two momenta related by a six-fold rotation.\\

We choose the parameters $t_1^A = -t_1^B = 1$, $t_2^A = -t_2^B = 0.5$, $t_1' =2.5$, $t_2' =0.5$. For $t_z' = 0.5 (\text{resp.}\;2)$ , the phase is trivial (resp. strong) and its Berry-phase spectrum is illustrated in Fig.\ \ref{fig:BZsWloops}(e) (resp.\ (f)). The surface modes are respectively illustrated in Fig.\ \ref{fig:C6Tcomposite}(a) and (b).

\section{Review of Wilson loops} \label{app:rvWilson}

As defined in Eq.\ (\ref{wloopdifferentiable}) for a non-contractible loop $l$, the matrix representation of holonomy is the Wilson loop $\W[l]$. In the $\noc$-dimensional basis of occupied bands, the generic Wilson loop  $\W \in U(\noc)$. Following standard convention, we define $U(\noc)$ ($SU(\noc)$) as the group of (special) unitary matrices in $\noc$ dimensions; similarly, $O(\noc)$ ($SO(\noc)$) as the group of (special) orthogonal matrices in $\noc$ dimensions. Since we focus on bands which derive from doublet irreps of $C_n+T$, $\noc$ is even. In Eq.\ (\ref{wloopdifferentiable}), the form of $\W[l]$ is applicable to a basis of wavefunctions which is differentiable in $\boldsymbol{k}$. By basis (or gauge), we mean a choice of a set of occupied Bloch wavefunctions $\{u_{i,\boldsymbol{k}}\}$, for each momentum along the loop $l$. Requiring a differentiable basis complicates any numerical computation. We are thus motivated in formulating a discretized expression of $\W[l]$; the discretization amounts to dividing $l$ into infinitesimally-separated momenta: $\{ \boldsymbol{k^{\sma{(0)}}}+\boldsymbol{G},\boldsymbol{k^{\sma{(N\text{-}1)}}},\boldsymbol{k^{\sma{(N\text{-}2)}}}, \ldots, \boldsymbol{k^{\sma{(2)}}}, \boldsymbol{k^{\sma{(1)}}},\boldsymbol{k^{\sma{(0)}}} \}$, with $N \rightarrow \infty$ and $\boldsymbol{G}$ a reciprocal lattice vector. Let us define $P(\boldsymbol{k}) = \sum_{i=1}^{\noc} \ket{u_{i,\boldsymbol{k}}}\bra{u_{i,\boldsymbol{k}}}$ as the projection to the occupied bands. $\W$ is expressed as path-ordered product of projections, sandwiched by tight-binding eigenfunctions at the base and end points\cite{AA2014}, i.e., it has matrix elements
\begin{align} \label{eq:wilsonprojection}
\W[l]_{ij} = \bra{u_{i,\boldsymbol{k^{\sma{(0)}}}+\boldsymbol{G}}} \; {\displaystyle \prod_{\alpha}^{\boldsymbol{k^{\sma{(0)}}}+\boldsymbol{G} \leftarrow \boldsymbol{k^{\sma{(0)}}}}}\, P(\boldsymbol{k^{\sma{(\alpha)}}})\; \ket{u_{j,\boldsymbol{k^{\sma{(0)}}}}}.
\end{align}
In this form, differentiability is manifestly not required. Implicit in the definition of $\W[l]$ is our choice of the so-called periodic gauge: $\ket{u_{i,\boldsymbol{k}+\boldsymbol{G}}} = \cald(\boldsymbol{G})^{\mo}\,\ket{u_{i,\boldsymbol{k}}}$, where $\cald(\boldsymbol{G})$ encodes the generic non-periodicity of our basis vectors  $\phi_{\boldsymbol{k},\alpha}$ under $\boldsymbol{k} \rightarrow \boldsymbol{k}+\boldsymbol{G}$, for a reciprocal lattice (RL) vector $\boldsymbol{G}$; cf. (\ref{basisvec}). This non-periodicity implies that the tight-binding Hamiltonian satisfies: $H(\boldsymbol{k}+\boldsymbol{G}) = \cald(\boldsymbol{G})^{\mo}\,H(\boldsymbol{k})\,\cald(\boldsymbol{G})$, where $\cald(\boldsymbol{G})$ is a unitary matrix with elements: $[\cald(\boldsymbol{G})]_{\ab} = \delta_{\ab}\,e^{iG\cdot \boldsymbol{r_{\alpha}}}$. In the proofs below, notation is greatly simplified if we assume that all orbitals within a unit cell lie on the same atom, i.e., $\cald(\boldsymbol{G})$ is the identity for any $\boldsymbol{G}$. The generalization to arbitrary spatial embeddings is straightforward, and we refer the interested reader to Ref.\ \onlinecite{AAchen}.\\  

Let us define the overlap matrix between two momenta by its matrix elements
\begin{align} \label{overlapmatrix}
\cals(\boldsymbol{k_1},\boldsymbol{k_2})_{ij} = \braket{u_{i,\boldsymbol{k_1}}}{u_{j,\boldsymbol{k_2}}}; \;\;\; i,j \in \{1,2, \ldots, n_o\}.
\end{align}
A third useful expression of the Wilson loop is as a path-ordered product of overlap matrices:
\begin{align} \label{wilsonoverlap}
\W[l]_{ij} = \sum_{r_1,r_2, \ldots r_{N-1}=1}^{\noc} \cals(\boldsymbol{k^{\sma{(0)}}}+\boldsymbol{G},\boldsymbol{k^{\sma{(N\text{-}1)}}})_{i,r_{N-1}}\,\cals(\boldsymbol{k^{\sma{(N\text{-}1)}}},\boldsymbol{k^{\sma{(N\text{-}2)}}})_{r_{N-1},r_{N-2}}\;\ldots\; \cals(\boldsymbol{k^{(2)}},\boldsymbol{k^{(1)}})_{r_2,r_1}\,\cals(\boldsymbol{k^{(1)}},\boldsymbol{k^{(0)}})_{r_1,j}
\end{align}
which follows directly from (\ref{eq:wilsonprojection}).

\end{widetext}

\begin{figure}
\centering
\includegraphics[width=8.6cm]{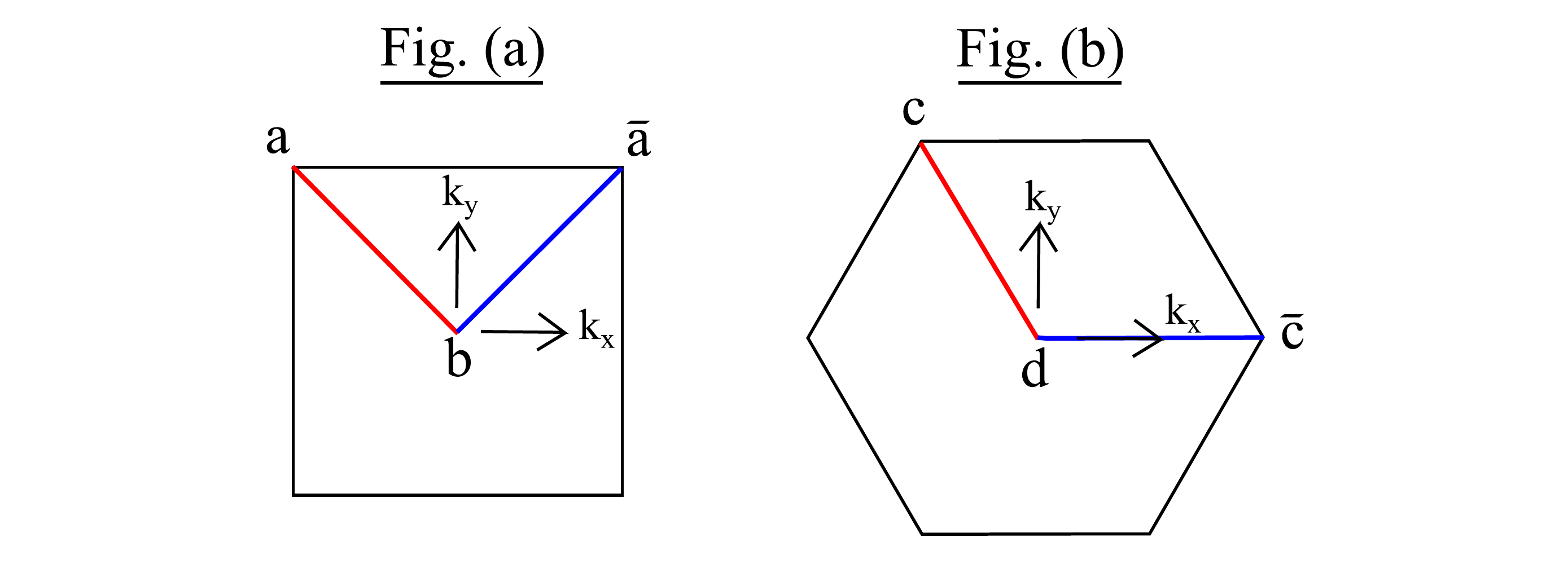}
\caption{(a) A constant-$k_z$ plane in the 3D Brillouin zone of a tetragonal lattice. $a,\bar{a}$ and $b$ are $C_4$-invariant momenta, with $a$ and $\bar{a}$ identified by a reciprocal lattice vector. (b)  A constant-$k_z$ plane in the 3D Brillouin zone of a hexagonal lattice. $c,\bar{c}$ and $d$ are $C_3$-invariant momenta, with $c$ and $\bar{c}$ identified by a reciprocal lattice vector.}\label{fig:bentwilson}
\end{figure}

The bulk of the appendix aims to characterize $\W[l_n({k}_z)]$, where $l_n({k}_z)$ denotes a bent loop in a plane of constant ${k}_z$; these loops are defined in the main text. It is convenient to express each of these loops as a product of two Wilson lines. In the case of $C_4+T$,
\begin{align} \label{c4wl}
\W[l_4(k_z)] = \W_{\sma{b \nearrow \bar{a}}}(k_z)\; \W_{\sma{a \searrow b}}(k_z) \in U(\noc),
\end{align}
where $ \W_{\sma{a \searrow b}}$ is an in-plane Wilson line that connects momenta $a=(-\pi,\pi,k_z)$ to $b=(0,0,k_z)$ along a diagonal in the upper-left quadrant (red line in Fig.\ \ref{fig:bentwilson}(a)); $\W_{\sma{b \nearrow \bar{a}}}$ is an in-plane Wilson line that connects $b=(0,0,k_z)$ to $\bar{a}=(\pi,\pi,k_z)$ along a diagonal in the upper-right quadrant (blue line in Fig.\ \ref{fig:bentwilson}(a)); the momenta $a$ and $\bar{a}$ are connected by a reciprocal lattice vector.  Analogously, we express
\begin{align} \label{c6wl}
\W[l_6(k_z)] = \W_{\sma{d \rightarrow \bar{c}}}(k_z)\;\W_{ \sma{{c} \searrow d}}(k_z) \in U(\noc),
\end{align}
where we define the $C_3$-invariant momenta: $c = (-2\pi/3,2\pi/\sqrt{3},k_z)$, $d = (0,0,k_z)$ and $\bar{c} = (4\pi/3,0,k_z)$; $c$ and $\bar{c}$ are connected by a reciprocal lattice vector. $\W_{ \sma{{c} \searrow d}}$ ($\W_{\sma{d \rightarrow \bar{c}}}$) is illustrated in red (blue) in Fig.\ \ref{fig:bentwilson}(b). Certain properties $\W[l_n(\bar{k}_z)]$ for $\bar{k}_z \in \{0,\pi\}$ are not shared by $\W[l_n({k}_z)]$ for general $k_z$. For example, the geometric interpretation of $\W$ as a special rotation only applies for $\bar{k}_z \in \{0,\pi\}$. The reader is thus advised to distinguish between $\bar{k}_z$ and $k_z$ in various contexts.

\section{Analytic properties of $\W[l_n(k_z)]\in U(\noc)$  for general $k_z$} \label{app:generalkz}

\subsection{Bounds on the eigenvalues of $\W[l_6(k_z)]\in U(\noc)$ for general $k_z$, as applied to the $C_6+T$ insulator} \label{eigenvalueinequality}

Consider the eigenvalues $\{\text{exp}[i\vartheta_j(k_z)]\}$ of a family of loops $\W[l_6(k_z)]\in U(\noc)$, which are defined in (\ref{c6wl}). $j \in \{1,2,\ldots,\noc\}$ labels the eigenvalue. We define $\vartheta_j$ as lying in the principal branch, i.e., $-\pi<\vartheta_j \leq \pi$. In this section we derive the following result: within this family of loops, $|\vartheta_j(k_z)| \leq 2\pi/3$ for all $j$ and $k_z$. \\

\noindent \emph{Proof}:  Following our definitions of $c,\bar{c},$ and $d$ in App.\ref{app:rvWilson}, the occupied bands $\{u_{i,\boldsymbol{k}}\}$ along the lines $c \searrow d$ and $d \rightarrow \bar{c}$ are related by $C_3$ symmetry, as illustrated in Fig.\ \ref{fig:bentwilson}(b). This motivates us to consider the matrix representation of $C_3$ in the basis of occupied doublet bands:
\begin{align} \label{C3matrix}
[\,B_{\sma{2\pi/3}}(\boldsymbol{k})\,]_{ij} = \bra{u_{i,R_{\sma{2\pi/3}}\boldsymbol{k}}}\;U_{\sma{2\pi/3}}\;\ket{u_{j,\boldsymbol{k}}},
\end{align}
where
\begin{align}
R_{\sma{2\pi/3}} \begin{pmatrix} k_x \\ k_y \\ k_z\end{pmatrix} = \begin{pmatrix} k_x\co 2\pi/3 - k_y\si 2\pi/3 \\ k_y\co 2\pi/3 + k_x\si 2\pi/3 \\ k_z \end{pmatrix},
\end{align}
and $U_{\sma{2\pi/3}}$ represents a rotation of $2\pi/3$ in the basis of \low orbitals. We define $\boldsymbol{k_i}$ as $C_3$-invariant momenta for which $R_{\sma{2\pi/3}}(\boldsymbol{k_i}) = \boldsymbol{k_i}$ up to reciprocal lattice vectors. The bands at $\boldsymbol{k_i}$ form doublets with $C_3$-eigenvalues exp$(\pm i 2\pi/3)$, thus a general form for $B(\boldsymbol{k_i})$ is 
\begin{align} \label{generalformC3sew}
B_{\sma{2\pi/3}}(\boldsymbol{k_i}) = F(\boldsymbol{k_i})\;e^{i \calr 2\pi/3}\;\dg{F(\boldsymbol{k_i})},
\end{align}
where $F(\boldsymbol{k_i}) \in U(\noc)$, $\calr= \bigoplus_{j=1}^m \sigma_3$ is a block-diagonal matrix and $\sigma_3$ is a Pauli matrix. Due to the $C_3$ symmetry relating bands along $c \searrow d$ and $d \rightarrow \bar{c}$, we deduce from (\ref{c6wl}) that
\begin{align} \label{generalwilson}
\W[l_6(k_z)] = \dg{B_{\sma{2\pi/3}}(c)}\,\dg{\W_{\sma{c \searrow d}}(k_z)}\,B_{\sma{2\pi/3}}(d)\,\W_{ \sma{c \searrow d}}(k_z).
\end{align}
Here, we have related $\W_{\sma{d \rightarrow \bar{c}}}$ to $\dg{\W_{ \sma{c \searrow d}}}$, where the Hermitian conjugation arises because rotating $c \searrow d$ by angle $2\pi/3$ produces $d \rightarrow \bar{c}$ with the reverse orientation. For more details of the algebraic steps leading to (\ref{generalwilson}), the interested reader may refer to an analogous calculation in Ref.\ \onlinecite{AA2014}.
Let us then define
\begin{align}
\tilde{R}(k_z) = \dg{F(c)}\,\dg{\W_{\sma{c \searrow d}}(k_z)}\,F(d)\,\calr\,\dg{F(d)}\,\W_{ \sma{c \searrow d}}(k_z)\,F(c).
\end{align}
Since $\dg{F(d)}\,\W_{ \sma{c \searrow d}}(k_z)\,F(c)$ is a product of unitary matrices, it is itself unitary. We then find that $\W[l_6(k_z)]$ is equivalent to $e^{-i \calr 2\pi/3}\,e^{i \tilde{R}(k_z) 2\pi/3}$, up to a unitary transformation. We denote this unitary equivalence by the symbol $\sim$: 
\begin{align}
\W[l_6(k_z)] \sim e^{-i \calr 2\pi/3}\,e^{i \tilde{R}(k_z) 2\pi/3} = e^{i \calr \pi/3}\;e^{-i \tilde{R}(k_z) \pi/3};
\end{align}
the last equality follows from $\calr^2= \tilde{R}^2=I$. We employ a theorem proven in Ref.\ \onlinecite{Nudelman,Agnihotri,Childs}, which sets a bound on the eigenvalues of a product of two unitary matrices $U_1$ and $U_2$, given the eigenvalues of each matrix. Let us denote the largest argument of the eigenvalue of a matrix $M$ as maxarg$[M]$. The theorem states: if maxarg$[U_1]$ + maxarg[$U_2] \leq \pi$, then maxarg$[U_1\,U_2] \leq $ maxarg$[U_1]$ + maxarg[$U_2]$. In our application, 
\begin{align}
\text{maxarg}[\;&e^{i \calr \pi/3}\;] = \text{maxarg}[\;e^{-i \tilde{R}(k_z) \pi/3}\;] = \pi/3 \notag \\
& \imp \text{maxarg}[\; \W[l_6(k_z)] \;] \leq 2\pi/3.
\end{align}
We point out that an analogous argument for $\W[l_4(k_z)]\in U(\noc)$, as applied to the $C_4+T$ insulator, leads to the trivial bound maxarg$[ \W[l_4(k_z)]] \leq \pi$.

\subsection{Proof that $\W[l_n(k_z)]\in U(\noc)$ has unit determinant, for general $k_z$} \label{app:unitdeterminant}

In the $C_6+T$ case, it follows directly from  (\ref{generalwilson}) and that $B_{\sma{2\pi/3}}(\boldsymbol{k_i})$ also has unit determinant, as made apparent in the form (\ref{generalformC3sew}).\\

The $C_4+T$ case proceeds in analogous fashion. We define the matrix representation of $C_4$ in the basis of occupied doublet bands:
\begin{align} \label{C4matrix}
[\,B_{\sma{\pi/2}}(\boldsymbol{k})\,]_{ij} = \bra{u_{i,R_{\sma{\pi/2}}\boldsymbol{k}}}\;U_{\sma{\pi/2}}\;\ket{u_{j,\boldsymbol{k}}};
\end{align}
$R_{\sma{\pi/2}}(k_x,k_y,k_z) = ( - k_y , k_x, k_z)$; $U_{\sma{\pi/2}}$ represents a rotation of $\pi/2$ in the basis of \low orbitals. We define $\boldsymbol{k_i}$ as $C_4$-invariant momenta for which $R_{\sma{\pi/2}}(\boldsymbol{k_i}) = \boldsymbol{k_i}$ up to reciprocal lattice vectors. The bands at $\boldsymbol{k_i}$ form doublets with $C_4$-eigenvalues $\pm i$, thus a general form for $B(\boldsymbol{k_i})$ is 
\begin{align} \label{generalformC4sew}
&B_{\sma{\pi/2}}(\boldsymbol{k_i}) = F(\boldsymbol{k_i})\;e^{i \calr \pi/2}\;\dg{F(\boldsymbol{k_i})}; \;\; F(\boldsymbol{k_i}) \in U(\noc),
\end{align}
where, again, $\calr= \bigoplus_{j=1}^m \sigma_3$. Following our definitions of $a,\bar{a},$ and $b$ in App.\ref{app:rvWilson}, the occupied bands $\{u_{i,\boldsymbol{k}}\}$ along the lines $b \nearrow \bar{a}$ and $a \searrow b$ are related by $C_4$ symmetry. Thus from (\ref{c4wl})\cite{AA2014},
\begin{align}  \label{genC4rela}
\W[l_4(k_z)] = \dg{B_{\sma{\pi/2}}(a)}\,\dg{\W_{\sma{a \searrow b}}(k_z)}\,B_{\sma{\pi/2}}(b)\,\W_{ \sma{a \searrow b}}(k_z).
\end{align}
It is clear from (\ref{generalformC4sew}) that $B_{\sma{\pi/2}}(\boldsymbol{k_i})$ has unit determinant, thus taking the determinant of (\ref{genC4rela}) produces the desired result.\\

\section{Analytic properties of $\W[l_n(\bar{k}_z)]\in SO(\noc)$  for specific $\bar{k}_z \in \{0,\pi\}$} \label{app:specifickz}

This appendix describes the bent Wilson loop $\W[l_n(\bar{k}_z)]$ in the high-symmetry planes $\bar{k}_z=0$ and $\pi$. In App.\ref{existrealgauge} and \ref{app:formsWL}, we show that in an appropriate basis, $\W$ is a special type of proper rotation. We describe the nature of this rotation in App.\ref{geometricargument}; here the discussion aims to develop a geometric intuition and is less technical. As a prelude to deriving the spectrum of $\W$ in App.\ref{app:spectrum}, we summarize in App.\ref{app:cartan} certain useful ideas in the representation theory of Lie groups.

\subsection{A real, periodic basis is found where $\W[l_n(\bar{k}_z)] \in SO(\noc)$} \label{existrealgauge}

\noindent \emph{Proof}: Let us define $C_2 \calt$ as the product of a two-fold rotation and a time reversal. In the planes of constant $\bar{k}_z \in \{0,\pi\}$, $C_2 {\cal T}$ maps a momentum to itself. We define the matrix representation of $C_2 {\cal T}$ in the basis of occupied bands as:  $[\,V_{\pi}(\boldsymbol{k})\,]_{ij} = \bra{u_{i,\boldsymbol{k}}}\;{U}_{\pi}\,T\;\ket{u_{j,\boldsymbol{k}}}$, with $T=Q\tilde{K}$ an antiunitary operator. Applying $(U_{\pi})^2=I$, $[U_{\pi},T] = 0$, and $Q=Q^t$, we find that ${U}_{\pi}\,Q$ (and hence $V_{\pi}$ also) is symmetric. A unitary, symmetric matrix can be written as $V_{\pi} = \calf\,D\,\calf^t$, where $\calf$ is a real orthogonal matrix and $D$ is a diagonal matrix with unimodular eigenvalues $\{\text{exp}(i\phi_j)\}$. Under $U(\noc)$ gauge transformations, $\ket{u_{i,\boldsymbol{k}}} \rightarrow \sum_{j}\ket{u_{j,\boldsymbol{k}}}\,\calj(\boldsymbol{k})_{ji}$, $V_{\pi}(\boldsymbol{k}) \rightarrow \calj(\boldsymbol{k})^t\,V_{\pi}(\boldsymbol{k})\,\calj(\boldsymbol{k})^*$. Choosing $\calj(\boldsymbol{k}) = \calf(\boldsymbol{k})$, we diagonalize $V_{\pi}$: ${U}_{\pi}\,T\,\ket{u_{j,\boldsymbol{k}}} = e^{i\phi_{j,\boldsymbol{k}}}\, \ket{u_{j,\boldsymbol{k}}}$. Note that the matrix representation of an antiunitary operator cannot generically be diagonalized; crucial to this diagonalization is that $V_{\pi}$ is symmetric, thus its eigenvectors can be chosen to be real. By a second transformation $\ket{u_{j,\boldsymbol{k}}} \rightarrow e^{i (\phi_{j,\boldsymbol{k}}-\pi)/2}\,\ket{u_{j,\boldsymbol{k}}}$, we arrive at: 
\begin{align} \label{gaugecond}
{U}_{\pi}\,T\,\ket{u_{j,\boldsymbol{k}}} = -\ket{u_{j,\boldsymbol{k}}}
\end{align}
for all $j \in \{1,2, \ldots, \noc\}$. Equivalently, this basis choice results in $V_{\pi}(\boldsymbol{k})=-I$, i.e., for each band $j$ and  momentum $\boldsymbol{k}$, $u_{j,\boldsymbol{k}}$ transforms in the \emph{real} vector representation of $C_2\calt$, e.g., a $p_x$ orbital. We thus call (\ref{gaugecond}) a real gauge. Now we show that $\W[l_n(\bar{k}_z)] \in SO(\noc)$ in a certain basis. The basis is defined by choosing a set of occupied Bloch wavefunctions $\{u_{i,\boldsymbol{k}}\}$, for each momentum along the loop $l_n(\bar{k}_z)$. Let us define $\boldsymbol{k^{(0)}}$ and $\boldsymbol{k^{(0)}}+\boldsymbol{G}$ as the base and end points of the loop, for some reciprocal lattice vector $\boldsymbol{G}$. \\

\noindent \emph{Definition}: In the real, periodic basis, (i) (\ref{gaugecond}) is satisfied for all $k$ in the loop $l_n(\bar{k}_z)$, and (ii) the basis is periodic in the sense of $\ket{u_{i,\boldsymbol{k^{(0)}}+\boldsymbol{G}}}=\ket{u_{i,\boldsymbol{k^{(0)}}}}$ for all $i \in \{1,2, \ldots, \noc\}$. \\

\noindent The second condition is implicit in the definition of the Wilson loop, and a generalization exists for nontrivial spatial embeddings of the orbitals\cite{AAchen}. One may verify that (i) and (ii) can be imposed consistently, though the resulting basis is not necessarily differentiable in $\boldsymbol{k}$; as explained in App.\ \ref{app:rvWilson}, differentiability is not required in the discrete formulation of $\W$. It should be noticed that (i) and (ii) does not fully specify the basis. While maintaining the reality condition (i), we are free to make orthogonal gauge transformations: $\ket{u_{i,\boldsymbol{k}}} \rightarrow \sum_{j}\ket{u_{j,\boldsymbol{k}}}\,\calu(\boldsymbol{k})_{ji}$ for $\calu(\boldsymbol{k}) \in O(\noc)$. The periodic condition (ii) is also maintained so long as $\calu(\boldsymbol{k^{(0)}})=\calu(\boldsymbol{k^{(0)}}+\boldsymbol{G})$.\\

It follows from (i) that the overlap matrix $\cals(\boldsymbol{k_1},\boldsymbol{k_2})$, as defined in (\ref{overlapmatrix}), is real for any $\boldsymbol{k_1}$ and $\boldsymbol{k_2}$ on the loop. Since $\W$ is expressible as a product of overlap matrices (as shown in (\ref{wilsonoverlap})), $\W=\W^* \in O(\noc)$.  Since $\W$ also has unit determinant, as proven in App.\ \ref{app:unitdeterminant}, $\W \in SO(\noc)$. By the same argument, one shows that the Wilson lines $\W_{\sma{b \nearrow \bar{a}}}, \W_{\sma{a \searrow b}}, \W_{\sma{d \rightarrow \bar{c}}}$ and $\W_{ \sma{{c} \searrow d}} \in O(\noc)$ in the real, periodic basis. These Wilson lines are defined in (\ref{c4wl}) and (\ref{c6wl}).\\


\subsection{Equivalent expressions of $\W[l_n(\bar{k}_z)]\in SO(\noc)$} \label{app:formsWL}

We work in a real, periodic basis where the Wilson loop $\W[l_n(\bar{k}_z)] \in SO(\noc)$ for even $\noc$, and the Wilson lines $\W_{\sma{b \nearrow \bar{a}}}, \W_{\sma{a \searrow b}}, \W_{\sma{d \rightarrow \bar{c}}}$ and $\W_{ \sma{{c} \searrow d}} \in O(\noc)$; cf. App.\ref{existrealgauge}. This gauge choice is analytically convenient; note that our final result, the spectrum of the Wilson loop, is gauge-invariant. We show in this Section that all possible $\W[l_n(\bar{k}_z)]$ are classified into two groups, which are distinguished by an index $\Gamma_n(\bar{k}_z) \in \{ 1,\mo\}$. Specifically, there exists a basis in which
\begin{align} \label{specialgauge}
& \W[l_n(\bar{k}_z)] \sim  Z^t\,e^{\minus i2\pi\,S/n}\,e^{i(1-\Gamma_n(\bar{k}_z))\, M_{1,2}\, 2\pi/n}\,Z\,e^{i2\pi\,S/n},
\end{align}
where $Z \in SO(\noc)$, and $\Gamma_n(\bar{k}_z) \in \{+1,-1\}$. Here, $S$ is defined
\begin{align} \label{defineS}
S = \sum_{j=1}^{\noc/2}\;M_{2j-1,2j}
\end{align}
if the dimension of the space is $\noc$, and $M_{a,b}$ are generators of rotations in the $a-b$ plane:
\begin{align} \label{defineMab}
[\,M_{a,b}\,]_{ij} = -i\,\delta_{a,i}\,\delta_{b,j} + i\,\delta_{a,j}\,\delta_{b,i},
\end{align}
for $ a,b \in \{1,2,\ldots, \noc\}.$ Due to the analogy with rotations in $\R^{\noc}$, we refer to a two-dimensional subspace as a `plane' in the space of occupied bands. Note that the presence of $M_{1,2}$ in (\ref{specialgauge}) does not imply that this particular plane is special. As we will shortly clarify, by a different choice of basis (i.e., a redefinition of $Z$), one may have selected $M_{3,4}$ instead of $M_{1,2}$, or $M_{5,6}$, etc. We are interested in determining the spectrum of $\W[l_n(\bar{k}_z)]$, which is identical to the spectrum of the $SO(\noc)$ matrix
\begin{align}
\W_n(\Gamma_n) = Z^t\,e^{\minus i2\pi\,S/n}\,e^{i(1-\Gamma_n)\, M_{1,2}\, 2\pi/n}\,Z\,e^{i2\pi\,S/n},
\end{align}
as follows simply from (\ref{specialgauge}). To simplify notation, we will often suppress the dependences of various quantities on $\bar{k}_z \in \{0,\pi\}$. \\

Before arriving at (\ref{specialgauge}), we first derive a canonical form of  the matrix representation of $\cnt$. In the basis of occupied doublet bands, the matrix is defined by
\begin{align} \label{sewingmatrix}
[\,V_{\sma{2\pi/n}}(\boldsymbol{k})\,]_{ij} = \bra{u_{i,-R_{\sma{2\pi/n}}\boldsymbol{k}}}\;U_{\sma{2\pi/n}} T\;\ket{u_{j,\boldsymbol{k}}}.
\end{align}
$R_{\sma{2\pi/n}}$ implements a rotation of $2\pi/n$ in momentum space: $R_{\sma{2\pi/n}}(k_x,k_y,k_z) = ( k_x \co 2\pi/n + k_y \si 2\pi/n, k_y \co 2\pi/n - k_x \si 2\pi/n, k_z)$; $U_{\sma{2\pi/n}}$ represents a rotation of $2\pi/n$ in the basis of \low orbitals; $T$ is the time-reversal operator. We define $\boldsymbol{k_i}$ as $C_n$-invariant momenta for which $R_{\sma{2\pi/n}}(\boldsymbol{k_i}) = -\boldsymbol{k_i}$ up to reciprocal lattice vectors. A basis may be found where $V_{\sma{2\pi/n}}(\boldsymbol{k_i})$ has the canonical form:
\begin{align} \label{canonicalform}
&V_{\sma{2\pi/n}}(\boldsymbol{k_i}) = E(\boldsymbol{k_i})\;e^{i 2\pi\,S /n}\;E(\boldsymbol{k_i})^t,
\end{align}
where $E(\boldsymbol{k_i}) \in O(\noc)$ and $S$ is defined in (\ref{defineS}). This canonical basis is both real and periodic, as defined in App.\ref{existrealgauge}. The derivation of (\ref{canonicalform}) from (\ref{sewingmatrix}) is shown in Ref.\ \onlinecite{fu2011}, for the $C_4+T$ insulator. We construct the canonical basis explicitly for the $C_6+T$ insulator.\\

\noindent \emph{Construction}: Since the Bloch Hamiltonian is three-fold symmetric at $\boldsymbol{k_i}$, $\{u_{n,\boldsymbol{k_i}}\}$ divide into $\noc/2$ pairs which transform in the doublet irrep of $C_3+T$; by assumption the one-dimensional irreps are absent. To be explicit, let us define in each doublet subspace the basis vectors $\ket{\omega}$ and $\ket{\omega^*} = U_{\sma{\pi/3}}\,T\,\ket{\omega}$, where $\ket{\omega}$ has $C_3$-eigenvalue $\omega=\text{exp}[i2\pi/3]$, and $\ket{\omega^*}$ has $\ct$-eigenvalue $\omega^*$. We form the orthogonal linear combinations: $\ket{1} = \alpha \ket{\omega} + \beta \ket{\omega^*}$ and $\ket{2} = - i \alpha \ket{\omega} +i \beta \ket{\omega^*}$. To satisfy (\ref{gaugecond}), i.e., ${U}_{\pi}\,T\,\ket{1} = - \ket{1}$, we set  $\alpha = \minus \beta^* \omega^*$. The reality condition (\ref{gaugecond}) implies that $[\,V_{\sma{\pi/3}}(\boldsymbol{k_i})\,]_{ij} = -\bra{u_{i,\boldsymbol{k_i}}}\;{U_{\sma{2\pi/3}}}^{\mo}\;\ket{u_{j,\boldsymbol{k_i}}}$ for $i,j$ restricted to this doublet subspace. The projection of ${U_{\sma{2\pi/3}}}^{\mo}$ onto the reduced basis $\{\ket{1},\ket{2}\}$ is exp$[\minus i 2\pi \sigma_2/3]$, where $\sigma_2$ is a Pauli matrix. This  construction can be independently obtained within each of the $\noc/2$ doublet subspaces. Thus we have found a canonical basis in which $\bra{u_{i,\boldsymbol{k_i}}}\;{U_{\sma{2\pi/3}}}^{\mo}\;\ket{u_{j,\boldsymbol{k_i}}}$ is a direct sum $\bigoplus_{j=1}^{\noc/2} \text{exp}[-i2\pi \sigma_2 /3]$, for $\noc$ occupied bands. Applying (\ref{defineS}) and $S^2=I$, we finally derive $V_{\sma{\pi/3}}(\boldsymbol{k_i}) = E(\boldsymbol{k_i})\;\big(-\text{exp}[-iS 2\pi/3]\big)\;E(\boldsymbol{k_i})^t =E(\boldsymbol{k_i})\;\text{exp}[{i S \pi/3}]\;E(\boldsymbol{k_i})^t$. Now we are ready to derive (\ref{specialgauge}); we separately tackle the $C_6+T$ and $C_4+T$ cases.


\subsubsection{Proof of (\ref{specialgauge}) for $\W[l_6(\bar{k}_z)] \in SO(\noc)$} \label{appsec:c6proof}

We recall the definitions of the $C_3$-invariant momenta: $c = (-2\pi/3,2\pi/\sqrt{3},k_z)$, $d = (0,0,k_z)$ and $\bar{c} = (4\pi/3,0,k_z)$, as illustrated in Fig.\ \ref{fig:bentwilson}-(b). In the planes of constant $\bar{k}_z$, the occupied bands $\{u_{i,k}\}$ along the lines $\bar{c} \searrow d$ and $d \rightarrow c$ are related by $C_6{\cal T}$ symmetry. Thus from (\ref{c6wl})\cite{AA2014},
\begin{align} \label{proofpart}
& \W[l_6(\bar{k}_z)] = \W_{\sma{d \rightarrow \bar{c}}}\,V_{\sma{\pi/3}}(d)^t\,{\W_{\sma{d \rightarrow \bar{c}}}}^t\,V_{\sma{\pi/3}}(c)
\end{align}
and $\W_{\sma{d \rightarrow \bar{c}}} \in O(\noc)$. Applying (\ref{canonicalform}), 
\begin{align} \label{ggy}
&\W[l_6(\bar{k}_z)] \notag \\
&\sim E(c)^t\;{\W_{\sma{ d \rightarrow \bar{c}}}}\;E(d)\;e^{-iS\pi/3}\;E(d)^t\;{\W_{\sma{ d \rightarrow \bar{c} }}}^t\;E(c)\;e^{iS \pi/3};
\end{align}
$\sim$ indicates an equivalence up to a unitary transformation. If det[$E(d)^t\,{\W_{\sma{ d \rightarrow \bar{c} }}}^t\,E(c)]=1$, we arrive at (\ref{specialgauge}) for $\Gamma_6=1$, with the identification $Z = E(d)^t\,{\W_{\sma{ d \rightarrow \bar{c} }}}^t\,E(c) \in SO(\noc)$. If det[$E(d)^t\;{\W_{\sma{ d \rightarrow \bar{c} }}}^t\;E(c)]=-1$, we employ the identity
\begin{align} \label{identity1}
e^{-i S \pi/3} = N_{2} \;e^{-i S \pi/3 + 2i M_{1,2} \pi/3}\;N_{2}^t,
\end{align}
where $N_{a}$ is a simple reflection with reflection axis in the direction $a$, i.e.,  $N_a$ has matrix elements 
\begin{align} \label{matrixelementsNa}
[\,N_{a}\,]_{ij} = \delta_{ij} - 2\,\delta_{a,i}\,\delta_{a,j}.
\end{align}
(\ref{identity1}) follows simply from the Clifford algebra of Pauli matrices, since $M_{1,2}$ acts like the Pauli matrix $\sigma_2$ in the one-two plane, and $N_2$ like $\sigma_3$. Inserting (\ref{identity1}) into (\ref{ggy}) we derive (\ref{specialgauge}) for $\Gamma_6=\mo$, with the identification $Z = N_2^t\,E(d)^t\,{\W_{\sma{ d \rightarrow \bar{c} }}}^t\,E(c) \in SO(\noc)$. More generally,
\begin{align} \label{generalidenref}
N_aM_{b,c} =\begin{cases} M_{b,c}N_a  & \ins{if} a \neq b, a \neq c, \\ -M_{b,c}N_a & \ins{if} a =b \;\text{or}\; a=c. \end{cases}
\end{align}
In the first case, $M$ and $N$ act in different subspaces; if they act in the same subspace, they anticommute. This general identity leads to
\begin{align} 
e^{-i S \pi/3} = N_{2a} \;e^{-i S \pi/3 + 2i M_{2a-1,2a} \pi/3}\;N_{2a}^t,
\end{align}
for $a \in \{1,2,\ldots, \noc/2\}$. If we had employed this identity with $a=2$, we would have derived (\ref{specialgauge}) with $M_{1,2}$ replaced by $M_{3,4}$, and $Z=N_4^t\,E(d)^t\,{\W_{\sma{ d \rightarrow \bar{c} }}}^t\,E(c)$. The spectrum of $\W$ does not depend on this choice of plane. 

\subsubsection{Proof of (\ref{specialgauge}) for $\W[l_4(\bar{k}_z)] \in SO(\noc)$}  \label{appsec:c4proof}

We recall the definitions the $C_4$-invariant momenta: $a=(-\pi,\pi,k_z)$, $b=(0,0,k_z)$, and $\bar{a}=(\pi,\pi,k_z)$, as illustrated in Fig.\ \ref{fig:bentwilson}-(a). In the planes $k_z=0$ and $\pi$, the occupied bands $\{u_{i,k}\}$ along the lines $b \nearrow \bar{a}$ and $a \searrow b$ are related by $C_4{\cal T}$ symmetry. Thus from (\ref{c4wl})\cite{AA2014},
\begin{align}
\W[l_4(\bar{k}_z)]=  \W_{\sma{b \nearrow \bar{a}}}\,{V_{\sma{\pi/2}}(b)}^t\;{\W_{\sma{b \nearrow \bar{a}}}}^t\;V_{\sma{\pi/2}}(a).
\end{align}
The steps leading to (\ref{specialgauge}) are similar to those in App.\ref{appsec:c6proof}.



\subsubsection{Alternative expression of $\W[l_n(\bar{k}_z)]\in SO(\noc)$, for $\noc=4m+2$ and integral $m$} 

For $\noc =4m+2$ and integral $m$, a useful expression is:
\begin{align} \label{so4np2WLform}
\W[l_n(\bar{k}_z)] \sim  X^t\,e^{- i 2\pi\,\Gamma_n(\bar{k}_z)\,S /n}\,X\,e^{i2\pi\,S/n} \in SO(4m+2),
\end{align}
with $X \in SO(4m+2)$ and $\Gamma_n(\bar{k}_z) \in \{+1,-1\}$.
Let us define the right-hand-side of this equation as
\begin{align}
\tilde{\W}_n(\Gamma_n) = X^t\,e^{- i 2\pi\,{\Gamma_n}\,S /n}\,X\,e^{i2\pi\,S/n} \in SO(4m+2).
\end{align}
(\ref{so4np2WLform}) follows directly from (\ref{specialgauge}) if $\Gamma_n=1$. If $\Gamma_n = -1$, we employ the identity
\begin{align} \label{idenV}
e^{- i2\pi\,S/n}\,e^{i\, M_{1,2}\,4\pi/n} = \calv\;e^{i2\pi\,S /n}\;\calv^t,  
\end{align}
where
\begin{align}
\calv =\calv^t = \prod_{a=2}^{2m+1}\,N_{2a} \in SO(4m+2)
\end{align}
is a product of an even number of simple reflections, as defined in (\ref{matrixelementsNa}). (\ref{idenV}) follows from applying the identity (\ref{generalidenref}) $2m$ times, as we now show:
\begin{align}
&\ldots N_8 N_6 N_4 e^{i2\pi\,S /n} N_4^tN_6^tN_8^t \ldots \notag \\
\eq \ldots N_8 N_6 e^{i2\pi\,S /n-i4\pi M_{3,4}/n} N_6^tN_8^t \ldots \notag \\
\eq \ldots N_8 e^{i2\pi\,S /n-i4\pi (M_{3,4}+M_{5,6})/n} N_8^t \ldots
\end{align}
Finally, we insert (\ref{idenV}) into (\ref{specialgauge}), thus obtaining (\ref{so4np2WLform}) with $X=\calv^tZ \in SO(4m+2)$.

\subsection{Geometric interpretation of $\W[l_n(\bar{k}_z)]\in SO(\noc)$ } \label{geometricargument}

In this Section we develop a geometric interpretation of $\W[l_n(\bar{k}_z)]\in SO(\noc)$ for $\noc=2m$, and its eigenspectrum $\{\text{exp}(i \vartheta)\}$. A rotation $R$ in $2m$ dimensions is described by $m$ invariant planes, and an angle of rotation in each plane; if all $m$ angles equal to $\theta$, such a rotation is called equiangular. Equivalently stated, an equiangular rotation acts as $e^{\pm i \sigma_2\theta}$ in each of the $m$ invariant planes, thus there exists a basis in which $R=e^{iJ\theta}$, where $J=\sum_{a=1}^{\noc/2} \eta_{2a-1,2a}M_{2a-1,2a}$ and $\eta_{2a-1,2a} \in \{+1,-1\}$; $M_{a,b}$ is the generator of rotation in the $a-b$ plane, as defined in (\ref{defineMab}). From this definition, note that $J=-J^t$, $J=\dg{J}$ and $J^2=I$. We have shown that $\W[l_n(\bar{k}_z)]$ is similar to (\ref{specialgauge}), which is a product of two equiangular rotations: 
\begin{align} 
\W_n(\Gamma_n) = R_2\bigg( \frac{2\pi}{n} \bigg)\,R_1\bigg( \frac{2\pi}{n} \bigg),
\end{align} 
where we define 
\begin{align}
&R_1\big( \phi \big) = e^{i \,S \,\phi}, \;\; R_2\big( \phi \big) = e^{i \,\tilde{Y} \,\phi}, \notag \\
&\tilde{Y} = Z^t \,[\,(1-\Gamma_n)M_{1,2}-S\,]\,Z,
\end{align}
for $Z \in SO(\noc)$. One may verify that 
\begin{align} \label{propertiesofequi}
S^2=&\tilde{Y}^2=I, \;\;\;S = \dg{S},\;\;\; \tilde{Y} = \dg{\tilde{Y}},\notag \\
& S=-S^t \;\;\;\text{ and}\;\;\; \tilde{Y}=-\tilde{Y}^t,
\end{align}
from the definitions of $M_{a,b}$ and $S$ in (\ref{defineMab}) and (\ref{defineS}). Each Berry phase is then interpreted as the net angle of rotation due to two equiangular rotations. \\

For any real vector $v$, it follows from (\ref{propertiesofequi}) that $iSv$ is also real and orthogonal to $v$. $v$ and $iSv$ span a plane ${\cal P}_1$ that is invariant under $R_1$; similarly, $\{v, i\tilde{Y}v\}$ span a plane ${\cal P}_2$ that is invariant under $R_2$. It follows from (\ref{propertiesofequi}) that the following norms are equal: $||iSv||=||i\tilde{Y}v||= ||v||$. Suppose there exists a vector $\bar{v}$ such that ${\cal P}_1$ and ${\cal P}_2$ coincide. Since both $iS\bar{v}$ and $i\tilde{Y}\bar{v}$ are real and have equal norms, this implies one of two cases: $S\bar{v} = \pm \tilde{Y}\bar{v}$. Let us define three orthogonal subspaces whose union forms the complete space.\\

\noi{a} If $S\bar{v} = - \tilde{Y}\bar{v}$, $R_1$ rotates  $w_{+}$  in a sense that is opposite to $R_2$. This is true of $w_{-}$ as well, hence both $w_{\pm}$ are invariant under $R_2 R_1$, i.e., each of $w_{\pm}$ has eigenvalue $1$. We define the invariant subspace $\calg_i$ as the kernel of $S+\tilde{Y}$.\\

\noi{b} If $S\bar{v} = \tilde{Y}\bar{v}$, $R_1$ rotates $w_{+}$ in the same sense as that for $R_2$, and this is true of $w_{-}$ as well. Then $w_{\pm}$ are maximally rotated under $R_2R_1$, i.e., $w_{\pm}$ have eigenvalues exp$({\pm i4\pi/n})$. We define the maximally-rotated subspace $\calg_m$ as the kernel of $S-\tilde{Y}$. The dimension of $\calg_m$ enters the weak index (\ref{2dz2inv}) as the quantity $d_n$.  \\

\noi{c} Since vectors in $\calg_m$ and $\calg_i$ have different eigenvalues under the same rotation, $\calg_m \cap \calg_i =0$. ${\cal G}^{\perp}$ is defined as the orthogonal complement to $\calg_i \cup \calg_m$. For each vector $v \in {\cal G}^{\perp}$, the plane of rotation under $R_1$ does not coincide with the plane of rotation under $R_2$. Then $v$ is neither invariant nor maximally rotated -- the net angle of rotation lies in the intermediate interval: $0<|\vartheta|<4\pi/n$. We had previously derived an identical bound in Sec. \ref{eigenvalueinequality} for a product of unitary matrices, which clearly also applies for a product of equiangular rotations.


\subsection{Cartan Decomposition of $SO(2m)$} \label{app:cartan}

Here we summarize certain ideas in the representation theory of Lie groups, which will aid in deriving the spectrum of $\W[l_n(\bar{k}_z)]$. The $SO(2m)$ group is generated by a semisimple Lie algebra $\call$, which can be decomposed into two subspaces: $\call = \calk + \tilde{P}$. These subspaces satisfy $ [\calk,\calk] \subseteq \calk, [\calk,\tilde{P}] \subseteq \tilde{P},$ and $[\tilde{P},\tilde{P}] \subseteq \calk$. Here, $[\calk,\calk] \subseteq \calk$ means: if $\tilde{k}_1,\tilde{k}_2 \in \calk$, then $[\tilde{k}_1,\tilde{k}_2] \in \calk$. $\calk$ is a subalgebra of $\call$, but $\tilde{P}$ is not. The Cartan subalgebra, as denoted by $\tilde{A}$, is defined as the maximal Abelian subalgebra that is contained in $\tilde{P}$; the dimension of $\tilde{A}$ is called the rank ($r$) of the decomposition. For an introduction to Cartan decompositions, see Ref.\ \onlinecite{Dagli} and references therein. Any matrix $Z \in SO(2m)$ may be expressed as
\begin{align} \label{cartandecompmatrix}
Z = e^{i\tilde{k}_1}\;\prod_{j=1}^r e^{i \Theta_j a_j}\; e^{i\tilde{k}_2} \in SO(2m)
\end{align}
for some real numbers $\{\Theta_1,\Theta_2,\ldots,\Theta_r\}$. Here, $\tilde{k}_1,\tilde{k}_2 \in \calk$ and $a_j \in \tilde{A}$; $SO(2)$ is a special case in which $\call=\calk$. Though we are not concerned with odd spatial dimensions in this paper, (\ref{cartandecompmatrix}) directly generalizes the Euler parametrization of a rotation in three dimensions. Just as the Euler decomposition is very useful in representing 3D rotations, we find that the representation (\ref{cartandecompmatrix}) reveals structure that is easily exploited, when deriving the spectrum of $\W[l_n(\bar{k}_z)]$. There are only two inequivalent types of Cartan decompositions of $SO(2m)$, for $m>1$\cite{Helgasonbook}. Their technical details are reported in the next two subsections. It is possible to skip these subsections on a first reading, and return to them when in need.

\subsubsection{Type-D-III Decomposition of $SO(2m)$; $m>1$} \label{Cartandecomp1}

For a type-D-III decomposition, $\calk$ is a $U(m)$ subalgebra of $SO(2m)$, and one choice of $\calk$ is spanned by $m^2$ vectors:
\begin{align}
 M_{2j-1,2j}&, \half \big(\;M_{2j-1,2k} + M_{2k-1,2j}\;\big),\notag \\
&\half \big(\;M_{2j-1,2k-1} + M_{2j,2k}\;\big),
\end{align}
for $j,k \in \{1,\ldots, m\}$ and $j\neq k$; the matrices $M_{a,b}$ are defined in (\ref{defineMab}). The $U(m)$ subalgebra contains a $U(1)$ subalgebra which is generated by $S$, as defined in (\ref{defineS}). $\tilde{P}$ is spanned by $m(m-1)$ vectors:
\begin{align}
\half \big(\;M_{2j-1,2k} - M_{2k-1,2j}\;\big), \half \big(\;M_{2j-1,2k-1} - M_{2j,2k}\;\big).
\end{align}
If $m$ is even, the rank $r=m/2$, and $\tilde{A}$ is spanned by $(\,M_{1+4j,4+4j} - M_{3+4j,2+4j}\,)$ for $j =0,1,\ldots, m/2-1$. If $m>1$ and is odd, $r=(m-1)/2$, and $\tilde{A}$ is spanned by $(\,M_{1+4j,4+4j} - M_{3+4j,2+4j}\,)$ for $j =0,1,\ldots, (m-1)/2-1$. 

\subsubsection{Type-BD-I Decomposition of $SO(2m)$; $m>1$} \label{BDI}

We perform a type-BD-I decomposition, where $\calk$ is a $SO(2) \times SO(2m-2)$ subalgebra of $SO(2m)$. One choice of $\calk$ is spanned by $M_{1,2}$ and all other $M_{a,b}$ such that $a,b \notin \{1,2\}$. The subspace  $\tilde{P}$ is spanned by $ M_{1,3},M_{1,4},\;\ldots\;,M_{1,2m}$ and $ M_{2,3},M_{2,4},\;\ldots\;,M_{2,2m}.$ The rank $r=2$, and $\tilde{A}$ is spanned by $M_{1,3}$ and $M_{2,4}$.

\subsection{Spectrum of $\W[l_n(\bar{k}_z)]\in SO(\noc)$,  for specific $\bar{k}_z \in \{0,\pi\}$} \label{app:spectrum}

Our aim is to prove Tab. \ref{C4spectrumtable}. We know from App.\ref{app:formsWL} that the set of all $\W[l_n(\bar{k}_z)]\in SO(\noc)$ fall into two classes, which are distinguished by an index $\Gamma_n \in \{1,\mo\}$. The proof is organized into five parts:  (i) In App.\ref{specSO2}, we solve for the spectrum of $\W[l_n(\bar{k}_z)]\in SO(2)$, i.e., with two occupied bands.  (ii) In App.\ref{SO4}, $SO(4)$. (iii) In App.\ref{4ntrivial}, $SO(4m)$ for $m \geq 2$ and $\Gamma_n=1$. (iv) In App.\ref{so4nplus2},  $SO(4m+2)$ for $m \geq 1$. (v) Finally in App.\ref{so4nnontrivial}, $SO(4m)$ for $m \geq 2$ and $\Gamma_n=-1$.

\subsubsection{Spectrum of $\W[l_n(\bar{k}_z)]\in SO(2)$} \label{specSO2}

Applying (\ref{specialgauge}) for two occupied bands,
\begin{align} \label{wilsonB}
 \W[l_n(\bar{k}_z)] \sim \W_n(\Gamma_n) \eq Z^t\;e^{-i\Gamma_n\, M_{1,2}\, 2\pi/n}\;Z\;e^{iM_{1,2}2\pi/n}\notag \\
\eq e^{i(1-\Gamma_n)\,M_{1,2}\,2\pi/n},
\end{align}
where $Z \in SO(2)$ and $M_{a,b}$ is defined in (\ref{defineMab}). The last equality follows from the commutivity of all $SO(2)$ matrices. If $\Gamma_n = 1$, the Wilson-loop spectrum is $\{1,1\}$; if $\Gamma_n = -1$, the spectrum is instead $\{\text{exp}(i4\pi/n),\text{exp}(\minus i4\pi/n) \}$. 

\subsubsection{Spectrum of $\W[l_n(\bar{k}_z)]\in SO(4)$} \label{SO4}

Applying (\ref{specialgauge}) for four occupied bands,
\begin{align} \label{so4wilsonloop}
 & \W[l_n(\bar{k}_z)] \sim \W_n(\Gamma_n) \notag \\
\eq  Z^t\;e^{-i(\,\Gamma_n\, M_{1,2} + M_{3,4}\,)\, 2\pi/n}\;Z\;e^{i(\,M_{1,2}+M_{3,4}\,)\,2\pi/n},
\end{align}
with $Z \in SO(4)$. The six generators of $SO(4)$ can be chosen as
\begin{align*}
& \cala_1 = \half \big( \;\tau_2\,\otimes\,\sigma_1\;\big) && \cala_2 = -\half \big( \;\tau_2\,\otimes\,\sigma_3\;\big) \notag \\
& \cala_3 = \half \big( \;\tau_0\,\otimes\,\sigma_2\;\big) && \calb_1 = -\half \big( \;\tau_1\,\otimes\,\sigma_2\;\big) \notag \\
& \calb_2 = -\half \big( \;\tau_2\,\otimes\,\sigma_0\;\big) && \calb_3 = \half \big( \;\tau_3\,\otimes\,\sigma_2\;\big),  
\end{align*}
where $\{\tau_i\}$ and $\{\sigma_i\}$ are Pauli matrices. Each of $\{\cala_i\}$ and $\{\calb_i\}$ generate an $SU(2)$ subalgebra of $SO(4)$, i.e., $[\,\cala_i,\cala_j\,]=i \epsilon_{ijk}\,\cala_k$, $[\,\calb_i,\calb_j\,]=i \epsilon_{ijk}\,\calb_k$ and $[\,\cala_i,\calb_j\,] = 0$. This is the well-known homomorphism: $SO(4) \sim SU(2) \times SU(2)$. In connection with the matrices in (\ref{so4wilsonloop}), we identify $M_{1,2}+M_{3,4} = 2\cala_3$ and $M_{1,2}-M_{3,4}=2\calb_3$. Let us derive the spectrum for $\Gamma_n= \pm 1$ separately.

\vspace{5mm}
\begin{center}
\emph{b.-(i)$\;\;\;\Gamma_n=1$}
\end{center}
\vspace{3mm}

A generic $SO(4)$ matrix can be written as $Z = e^{-i \boldsymbol{a} \cdot \boldsymbol{\cala}}\,e^{-i \boldsymbol{b} \cdot \boldsymbol{\calb}}$ for some coefficients $\{a_1,a_2,a_3,b_1,b_2,b_3\}$. Here, we treat $\boldsymbol{a} = (a_1,a_2,a_3)$ as a three-vector with norm $||\boldsymbol{a}|| = (\,a_1^2 + a_2^2 +a_3^2\,)^{1/2},$ and $\boldsymbol{a} \cdot \boldsymbol{\cala} = \sum_{i=1}^3 a_i \cala_i$. Since functions of $\boldsymbol{\cala}$ and $\boldsymbol{\calb}$ commute,
\begin{align} \label{so4trivial}
 &\W_n(1) = Z^t\;e^{-i\,\cala_3\, 4\pi/n}\;Z\;e^{i\,\cala_3\,4\pi/n} \notag \\
\eq e^{i \boldsymbol{a} \cdot \boldsymbol{\cala}}\;e^{-i\,\cala_3\, 4\pi/n}\;e^{-i \boldsymbol{a} \cdot \boldsymbol{\cala}}\;e^{i\,\cala_3\,4\pi/n} \notag \\
\eq e^{-i \boldsymbol{a}' \cdot \boldsymbol{\cala}},
\end{align}
for some coefficients $\{a'_1,a'_2,a'_3\}$. The last equality follows from the closure property of $SU(2)$. In a basis that diagonalizes both $\cala_3$ and $\calb_3$, (\ref{so4trivial}) is expressible as the direct product $e^{-i \boldsymbol{a}' \cdot \boldsymbol{\sigma}/2} \otimes I$. In this form, we identify the eigenvalues as 
\begin{align}
& \text{eig}[\; \W_n(1)\;] = \text{eig}[\;e^{-i \boldsymbol{a}' \cdot \sigma /2} \otimes I\;] \notag \\
\eq \text{eig}[\; \begin{pmatrix} \lambda && 0 \\ 0 && \lambda^* \end{pmatrix} \otimes \begin{pmatrix} 1 && 0 \\ 0 && 1 \end{pmatrix}\;] \notag \\
\eq \{ \lambda, \lambda, \lambda^*, \lambda^*\}. 
\end{align}
Here, $\lambda$ is unimodular with phase $||\boldsymbol{a}'||/2$, and  we introduce $\text{eig}[\,\calo\,]$ to mean the spectrum of $\calo$.

\vspace{5mm}
\begin{center}
\emph{b.-(ii) $\;\;\;\Gamma_n=-1$}
\end{center}
\vspace{3mm}
\noindent  From (\ref{so4wilsonloop}), we express $Z = e^{-i \boldsymbol{a} \cdot \boldsymbol{\cala}}\,e^{-i \boldsymbol{b} \cdot \boldsymbol{\calb}}$. Since functions of $\boldsymbol{\cala}$ and $\boldsymbol{\calb}$ commute,
\begin{align} \label{so4nontrivial}
 &\W_n(\text{-}1) = Z^t\;e^{i\,\calb_3\, 4\pi/n}\;Z\;e^{i\,\cala_3\,4\pi/n} \notag \\
\eq  e^{i \boldsymbol{b} \cdot \boldsymbol{\calb}}\;e^{i\,\calb_3\, 4\pi/n}\;e^{-i \boldsymbol{b} \cdot \boldsymbol{\calb}}\;e^{i\,\cala_3\,4\pi/n} \notag \\
\eq e^{-i \boldsymbol{b}' \cdot \boldsymbol{\calb}}\;e^{i\,\cala_3\,4\pi/n},
\end{align}
for some coefficients $\{b'_1,b'_2,b'_3\}$. The last equality follows from the closure property of $SU(2)$. Since $\text{exp}(-i \boldsymbol{b}' \cdot \boldsymbol{\calb})$ and $\text{exp}(i\,\calb_3\, 4\pi/n)$ are related by a unitary tranformation, $||\boldsymbol{b}'|| = 4\pi/n$. In a basis that diagonalizes both $\cala_3$ and $\calb_3$, we express (\ref{so4nontrivial}) as the direct product $e^{i\sigma_3 2\pi/n} \otimes e^{-i \boldsymbol{b}' \cdot \boldsymbol{\sigma}/2} $. Then,
\begin{align}
& \text{eig}[\; \W_n(\text{-}1)\;] = \text{eig}[\;e^{i\sigma_3 2\pi/n} \otimes e^{-i \boldsymbol{b}' \cdot \boldsymbol{\sigma}/2}\;] \notag \\
\eq \text{eig}[\; \begin{pmatrix} e^{i2\pi/n} && 0 \\ 0 && e^{-i2\pi/n} \end{pmatrix} \otimes \begin{pmatrix} e^{i2\pi/n} && 0 \\ 0 && e^{-i2\pi/n} \end{pmatrix}\;] \notag \\
\eq \{ 1, 1, e^{i4\pi/n}, e^{-i4\pi/n}\}.
\end{align}

\subsubsection{Spectrum of $\W[l_n(\bar{k}_z)]\in SO(4m)$; $m \geq 2$; $\Gamma_n=1$} \label{4ntrivial}

We employ the form of $\W[l_n(\bar{k}_z)]\in SO(4m)$ in (\ref{specialgauge}), for $\Gamma_n=1$. Following App.\ref{Cartandecomp1}, we perform a type-D-III Cartan decomposition of the matrix $Z$ in (\ref{specialgauge}), i.e., we express $Z$ in the form (\ref{cartandecompmatrix}) for some coefficients $\{\Theta_1,\Theta_2,\ldots,\Theta_r\}$; the rank of the decomposition $r=m$; $\{\tilde{k}_1,\tilde{k}_2\}$ belong to a $U(2m)$ subalgebra (${\calk}$) of $SO(4m)$. The Cartan subalgebra is spanned by 
\begin{align} \label{spanner}
a_j = \oneover{2} \big(\,M_{4j-3,4j} - M_{4j-1,4j-2}\,)
\end{align} 
for $j \in \{1,2,\ldots,m\}$. As defined in (\ref{defineS}), the matrix $S$ of (\ref{specialgauge}) generates a $U(1)$ subalgebra in ${\calk}$, i.e., it commutes with all elements in the $U(2m)$ subgroup generated by $\calk$. Hence, (\ref{specialgauge}) simplifies to
\begin{align} \label{hehehe}
 \W_n(1) \sim \prod_{j=1}^m\;e^{-i\Theta_j a_j} \; e^{-i2\pi\,S/n} \;\prod_{u=1}^m\;e^{i\Theta_u a_u}\;e^{i2\pi\,S/n}
 \end{align}
 in some basis. Now insert the identity: 
 \begin{align}
 e^{i2\pi\,S/n} =  \prod_{j=1}^m e^{i  (\,M_{4j-3,4j-2} + M_{4j-1,4j}\,)\,2\pi/n},
 \end{align}
which follows simply from (\ref{defineS}). Let us denote the basis vectors by $\{v_i | i \in \{1,2,\ldots, 4m\} \,\}$. We define $\caln^{(j)}$ as the 4D subspace spanned by  $v_{4j-3},v_{4j-2},v_{4j-1}$ and $v_{4j}$. Since $a_j$ and $M_{4j-3,4j-2} + M_{4j-1,4j}$ act only in $\caln^{(j)}$, we may arrange (\ref{hehehe}) as
\begin{align} \label{scommutes}
 \W_n(1) \sim \prod_{j=1}^m &\bigg(\;e^{-i\Theta_j a_j}\;e^{-i (\,M_{4j-3,4j-2} + M_{4j-1,4j}\,) \,2\pi/n}\notag \\
&\times\;e^{i\Theta_j a_j}\;e^{i  (\,M_{4j-3,4j-2} + M_{4j-1,4j}\,)\,2\pi/n}\;\bigg) \notag \\
\eq \bigoplus_{j=1}^m {\cal Q}^{(j)}.
\end{align}
Equivalently, $ \W_n(1)$ diagonalizes into $m$ blocks  -- each block ${\cal Q}^{(j)} \in SO(4)$ acts in the subspace of $\caln^{(j)}$. We proceed to determine the eigenvalues within each block ${\cal Q}^{(j)}$. First, we show that $a_j$ and $M_{4j-3,4j-2} + M_{4j-1,4j}$ generate an $SU(2)$ subalgebra in $\caln^{(j)}$. The six generators of $SO(4)$ in each block may be chosen as
\begin{align}
& \cala^{(j)}_1= a_j= \half \big(\;M_{4j-3,4j} - M_{4j-1,4j-2}\;\big)\notag \\
&  \cala^{(j)}_2= \half \big(\;M_{4j-2,4j} - M_{4j-3,4j-1} \;\big)  \notag \\
& \cala^{(j)}_3= \half\big(\;M_{4j-3,4j-2} + M_{4j-1,4j}\;\big) \notag \\
& \calb^{(j)}_1= -\half \big(\;M_{4j-3,4j} + M_{4j-1,4j-2}\;\big) \notag \\
& \calb^{(j)}_2= -\half \big(\;M_{4j-2,4j} + M_{4j-3,4j-1} \;\big) \notag \\
& \calb^{(j)}_3= \half\big(\;M_{4j-3,4j-2} - M_{4j-1,4j}\;\big). 
\end{align}
Each of $\{\cala^{(j)}_i\}$ and $\{\calb^{(j)}_i\}$ generate an $SU(2)$ subalgebra of $SO(4)$, i.e., they satisfy $[\;\cala^{(j)}_l,\cala^{(j)}_m\;] = i \,\pdg{\epsilon}_{lmn}\,\cala^{(j)}_n$, $ [\;\calb^{(j)}_l,\calb^{(j)}_m\;] = i \,\pdg{\epsilon}_{lmn}\,\calb^{(j)}_n$, and $[\cala^{(j)}_m, \calb^{(j)}_n]=0$. Thus, we identify ${\cal Q}^{(j)}$ as an element in the $SU(2)$ subgroup generated by $\{\cala^{(j)}_i\}$. The spectrum in each of ${\cal Q}^{(j)}$ has been solved in App.\ref{SO4} -- the full spectrum comprises $m$ sets of complex-conjugate quartets:
\begin{align}
\text{eig}[\; \W_n(1)\;] = \{ \;\lambda_1,\lambda_1,\lambda_1^*,\lambda_1^*, \;\ldots\;,\lambda_m,\lambda_m,\lambda_m^*,\lambda_m^*\;\}.
\end{align}

\subsubsection{Spectrum of $\W[l_n(\bar{k}_z)]\in SO(4m+2)$, for $m \geq 1$} \label{so4nplus2}

We employ the form of $\W[l_n(\bar{k}_z)]\in SO(4m+2)$ in (\ref{so4np2WLform}). Following App.\ref{Cartandecomp1}, we perform a type-D-III Cartan decomposition of the matrix $X$ in (\ref{so4np2WLform}), i.e., we express $X$ in the form (\ref{cartandecompmatrix}) for some $\{\Theta_1,\Theta_2,\ldots, \Theta_r\}$;  the rank of the decomposition $r=m$; $\{\tilde{k}_1,\tilde{k}_2\}$ belong to a $U(2m+1)$ subalgebra (${\calk}$) of $SO(4m+2)$. The Cartan subalgebra is spanned by (\ref{spanner}). As defined in (\ref{defineS}), the matrix $S$ in (\ref{so4np2WLform}) commutes with all elements generated by $\calk$, hence (\ref{so4np2WLform}) simplifies to
\begin{align} \label{inthebasisof}
\tilde{\W}_n(\Gamma_n) \sim & \;\prod_{j=1}^m\;e^{-i\Theta_j a_j}\;e^{-i2\pi\, \Gamma_n\,S /n}\;\prod_{l=1}^m\;e^{i\Theta_l a_l}\;e^{i2\pi\,S/n}.
\end{align}
Let us denote the basis vectors by $\{v_i | i \in \{1,2,\ldots, 4m+2\} \,\}$, and the plane spanned by $v_i$ and $v_j$ as ${\cal P}_{ij}$. We consider the cases $\Gamma_n= \pm 1$ separately.

\vspace{5mm}
\begin{center}
\emph{d.-(i) $\;\;\;\Gamma_n=1$}
\end{center}
\vspace{3mm}

\noindent In the basis of (\ref{inthebasisof}), any vector  $\bar{v} \in {\cal P}_{4m+1,4m+2}$ is invariant under $ \tilde{\W}_n(1)$. Thus there are at least two unit eigenvalues in the spectrum of $\tilde{\W}_n(1)$. \\

\noindent \emph{Proof}: As defined in (\ref{spanner}), $a_j$ generates rotations in the planes ${\cal P}_{4j-3,4j}$ and ${\cal P}_{4j-1,4j-2}$, thus any vector in ${\cal P}_{4m+1,4m+2}$ is invariant under the rotation $\prod_{j=1}^m\,\text{exp}(i\Theta_j a_j)$. Moreover, any rotation of $\bar{v}$  that is induced by exp$({iS2\pi/n})$ is subsequently negated by exp$({\text{-}iS2\pi/n})$.\\

\noindent Within the orthogonal complement of ${\cal P}_{4m+1,4m+2}$, each $a_j$ acts in $\caln^{(j)}$, defined as the 4D subspace spanned by  $v_{4j-3},v_{4j-2},v_{4j-1}$ and $v_{4j}$. Therefore, $\tilde{\W}_n(1)$ diagonalizes into (i) two one-by-one blocks with unit element, in the plane ${\cal P}_{4m+1,4m+2}$, and (ii) $m$ four-by-four blocks, each labelled by ${\cal Q}^{(j)} \in SO(4)$: $ \tilde{\W}_n(1) = I_{2 \times 2} \oplus \bigoplus_{j=1}^m {\cal Q}^{(j)}.$ The spectrum in each of ${\cal Q}^{(j)}$ has been solved in App.\ref{SO4} -- the full spectrum is
\begin{align} \label{fullspec3}
&\text{eig}[\; \tilde{\W}_n(1)\;] \notag \\
\eq \{ \;1,\,1,\,\lambda_1,\lambda_1,\lambda_1^*,\lambda_1^*, \;\ldots\;,\lambda_m,\lambda_m,\lambda_m^*,\lambda_m^*\;\}.
\end{align}

\vspace{5mm}
\begin{center}
\emph{d.-(ii) $\;\;\;\Gamma_n=-1$}
\end{center}
\vspace{3mm}

A vector  $\bar{v} \in {\cal P}_{4m+1,4m+2}$ is (i) invariant under $\prod_{j=1}^m\,\text{exp}(i\Theta_j a_j)$, but (ii) is rotated by $4\pi/n$ due to a double application of exp$(i 2\pi \,S/n)$. The remainder of the proof is similar to the case of $\Gamma_n=+1$. We conclude that the full spectrum is
\begin{align} \label{4np2spec}
 \text{eig}[\; \tilde{\W}_n(\text{-}1)\;] = \{& \;e^{i4\pi/n},\,e^{{-i}4\pi/n},\,\lambda_1,\lambda_1,\lambda_1^*,\lambda_1^*,\notag \\
& \;\ldots\;,\lambda_m,\lambda_m,\lambda_m^*,\lambda_m^*\;\}.
\end{align}

\subsubsection{Spectrum of $\W[l_n(\bar{k}_z)]\in SO(4m)$, for $m \geq 2$ and $\Gamma_n=-1$} \label{so4nnontrivial}

We employ the form of $\W[l_n(\bar{k}_z)]\in SO(4m)$ in (\ref{specialgauge}), for $\Gamma_n=-1$, i.e., a basis is found where $\W[l_n(\bar{k}_z)] \sim \W_n(\text{-}1)$. Some useful notations are defined in App.\ref{geometricargument}:  $R_1(\phi) = \text{exp}({i  S \phi})$ and  $R_2(\phi) = \text{exp}(i \tilde{Y}\phi)$; $Y =2M_{1,2}-S$ and $\tilde{Y} = Z^t \,Y\, Z$ for $Z \in SO(4m)$. $M_{ab}$ and $S$ are defined in (\ref{defineMab}) and (\ref{defineS}). We first solve for the spectrum of $\W_4(\text{-}1)\in SO(4m)$.

\vspace{5mm}
\begin{center}
\emph{e.-(i) \hspace{2mm} Spectrum of $\W_4(\text{-}1)\in SO(4m)$, for $m \geq 2$}
\end{center}
\vspace{3mm}

For the $C_4+T$ insulator, $\W_4(\text{-}1) = R_2( \frac{\pi}{2} )\,R_1( \frac{\pi}{2} ) = i\,\tilde{Y}\,i\,S.$ We first derive two general properties of its spectrum:\\

\noi{i} Since $\W_4(\text{-}1)$ is a proper rotation in even dimensions, its spectrum comprises of complex-conjugate pairs. \\

\noi{ii} Moreover, there exists an antiunitary operator $\calo = iS \tilde{K}$, where $\tilde{K}$ implements complex-conjugation, such that
\begin{align} \label{kramer}
\calo \;\W_4(\text{-}1)  \;\calo^{\mo} =  i\,S\,i\,\tilde{Y} = {\W_4(\text{-}1)}^t={\W_4(\text{-}1)}^{\mo}.
\end{align}
Here, we applied that $S$ and $\tilde{Y}$ are each imaginary and skew-symmetric; cf. (\ref{propertiesofequi}). Since $\calo^2 = -I$, (\ref{kramer}) implies that each eigenvalue of $\W_4(\text{-}1)$ is doubly-degenerate\cite{AA2014}. \\

Following App.\ref{BDI}, we perform a type-BD-I Cartan decomposition of $Z$ in $\tilde{Y} =  Z^t\,Y\,Z$, i.e., we express $Z$ in the form (\ref{cartandecompmatrix}) for some $\{\Theta_1,\Theta_2\}$; the rank of the decomposition $r=2$; $\{\tilde{k}_1,\tilde{k}_2\}$ belong to a $SO(2) \times SO(4m-2)$ subalgebra (${\calk}$) of $SO(4m)$. The Cartan subalgebra is spanned by the generators $a_1 = M_{1,3}$ and  $a_2=M_{2,4}$, where $[a_1,a_2]=0$. Our strategy is to derive the spectrum for the special values $\Theta_1=\Theta_2=0$, and then prove that the spectrum for arbitrary $(\Theta_1,\Theta_2)$ has essentially the same structure. If $\Theta_1=\Theta_2=0$, 
\begin{align}
\W_4(\text{-}1) = e^{-i\tilde{k}_3}\,i\,Y\,e^{i\tilde{k}_3}\,i\,S
\end{align}
where $\text{exp}(i\tilde{k}_3) = \text{exp}({i\tilde{k}_1})\,\text{exp}(i\tilde{k}_2)$ for some $\tilde{k}_3 \in \calk$, due to the closure property of the subgroup generated by $\calk$. Let us define $\calp_{12}$ as the plane spanned by basis vector $v_1$ and $v_2$. Since none of $\calk$, $Y$ and $S$ couple $\calp_{12}$ to its orthogonal complement, $\W_4(\text{-}1)$ diagonalizes into two blocks, with dimensions $2$ and $(4m-2)$. The two-dimensional block has the form (\ref{wilsonB}) for $n=4$ and $\Gamma_4=-1$, so we may apply the results derived in App.\ref{specSO2} -- the spectrum in this block is $\{-1,-1\}$. The $(4m-2)$-dimensional block has the form (\ref{so4np2WLform}) for $n=4$ and $\Gamma_4=+1$; take care that the integer $m$ in this Section differs from that of (\ref{so4np2WLform}) by unity. We then apply the results derived in App.\ref{so4nplus2} for $\Gamma_4=+1$ -- the spectrum in this block is $\{ 1,1,\lambda_1,\lambda_1,\lambda_1^*,\lambda_1^*, \;\ldots\;,\lambda_{m-1},\lambda_{m-1},\lambda_{m-1}^*,\lambda_{m-1}^*\,\}.$ The full spectrum for $\Theta_1=\Theta_2=0$ is thus
\begin{align} \label{fullspec}
\text{eig}[\;\W_4(\text{-}1)\;] = \{ &\;-1,\,-1,\,1,\,1,\,\lambda_1,\lambda_1,\lambda_1^*,\lambda_1^*, \;\ldots\;,\notag \\
 &\lambda_{m-1},\lambda_{m-1},\lambda_{m-1}^*,\lambda_{m-1}^*\;\}.
\end{align}
Since $Z$ is analytic in both $\Theta_1$ and $\Theta_2$, so is the operator $\W_4(\text{-}1)$. This implies that the eigenvalues of $\W_4(\text{-}1)$ are continuous functions of $\Theta_1$ and $\Theta_2$\cite{katobook,knoppbook}. Thus we consider interpolating between $(\Theta_1,\Theta_2)=(0,0)$ to any values which can be nonzero. We note that the properties (i) and (ii) of the above discussion were derived for the most general form of the matrix $Z$, i.e., they are applicable for any value of $\Theta_1$ and $\Theta_2$. Together, (i) and (ii) imply that any eigenvalue that is not $\pm 1$ must belong to a complex-conjugate quartet: $\{\lambda,\lambda,\lambda^*,\lambda^*\}$ -- this is true throughout the interpolation. We have derived that the spectrum at $\Theta_1=\Theta_2=0$ comprises two eigenvalues at $+1$ and the two eigenvalues at $\mo$; in the interpolation, these four eigenvalues cannot combine to form a complex-conjugate quartet in any continuous fashion, hence they are invariants of the interpolation. Thus we have shown that spectrum of $\W_4(\text{-}1)$ for any finite $(\Theta_1,\Theta_2)$ has the same form as that in (\ref{fullspec}). In particular, we have shown that there are at least $2$ eigenvalues of $+1$; it is possible that one or more of the complex-conjugate quartets lie at $+1$ as well, hence the total number is more generally $2+4t$, for  $t$ a non-negative integer less than $m$. Similarly, the number of $-1$ eigenvalues might be $2+4u$, for $u$ a non-negative integer less than $m$; $u+t<m$. We proceed to evaluate the spectrum of $\W_6(\text{-}1)\in SO(4m)$ for the $C_6+T$ insulator.



\vspace{5mm}
\begin{center}	
\emph{e.-(ii) \hspace{2mm} Spectrum of $\W_6(\text{-}1)\in SO(4m)$, for $m \geq 2$}
\end{center}
\vspace{3mm}

\textbf{Lemma 1}: the spectrum of $\W_6(\text{-}1)$ consists of: (i) $2+4u$ number of $+1$ eigenvalues, and (ii) $1+2t$ pairs of exp($\pm i 2\pi/3)$, for some non-negative integers $u$ and $t$ that satisfy $u+t<m$.\\

\noindent \emph{Proof}: Though we are interested in the spectrum of $\W_6(\text{-}1) = R_2( \frac{\pi}{3} )\,R_1( \frac{\pi}{3} )$, let us first consider the more general problem of 
\begin{align}
\W(\phi) = R_2( \phi )\,R_1( \phi );\;\;\; 2|\phi| \leq \pi. 
\end{align}
As introduced in App.\ref{geometricargument}, $\R^{4m}$ may be divided into three orthogonal subspaces:
\begin{align} \label{decomp}
&\calg_i = \text{ker}[\,S + \tilde{Y}\,],\;\;\;\calg_m = \text{ker}[\,S - \tilde{Y}\,], \notag \\
&\;\;\;\text{and}\;\;\; \calg^{\perp} =  \R^{4m} \backslash (\,\calg_i \cup \calg_m\,).
\end{align} 
Here, ker$[\calo]$ means the kernel of the operator $\calo$. An eigenvector of $\W(\phi)$ in the invariant subspace $\calg_i$ has unit eigenvalue; in the maximally-rotated subspace $\calg_m$, the eigenvalues appear in pairs of exp$(\pm i2\phi)$.  The decomposition (\ref{decomp}) is manifestly independent of $\phi$. This suggests the following strategy: evaluate $\calg_i$ and $\calg_m$ for $\W(\phi)$ with $\phi \neq \pi/3$;  these subspaces must be identical for the matrix $\W(\pi/3) = \W_6(\text{-}1)$. Conveniently, we apply the result (\ref{fullspec}), for $\W_4(\text{-}1)= \W(\pi/2)$. In the case of $\phi=\pi/2$, we have derived that there are $2+4t$ ($2+4u$) number of $+1$ ($-1$) eigenvalues, for some non-negative integers $\{u,t\}$ which satisfy $u+t < m$. It follows that the dimension of $\calg_i$, denoted by dim[$\calg_i]$, equals $2+4t$, and dim[$\calg_m]=2+4u$. This proves Lemma 1.\\

We proceed to evaluate the spectrum of $\W_6(\text{-}1)$ in the subspace ${\cal G}^{\perp}$, in the case that $\text{dim}[\calg^{\perp}]=4(m-1-t-u)$ is nonzero. \textbf{Lemma 2}: the spectrum comprises $\text{dim}[\calg^{\perp}]/4$ sets of complex-conjugate quartets: $\{\lambda_i,\lambda_i,\lambda_i^*,\lambda_i^*\}$.\\ 

\noindent \emph{Proof}: Let us show that $\calg = \calg_m \cup \calg_i$ is spanned by the simultaneous eigenvectors of $\tilde{Y}$ and $S$. For any $v \in \calg_m$, the orthogonal linear combinations $w_{\pm} = v \pm Sv \in \calg_m$ satisfy $S w_{\pm} = \tilde{Y} w_{\pm}= \pm w_{\pm}$. Similarly, for any $y \in \calg_i$, the orthogonal linear combinations $z_{\pm} = y \pm Sy \in \calg_i$ satisfy $S z_{\pm} = - \tilde{Y} z_{\pm}= \pm z_{\pm}$. Denoting the projection to a subspace $\calc$ by $P(\calc)$, it follows that 
\begin{align} \label{reducedW}
 & [\,P(\calg),S\,] = [\,P(\calg),\tilde{Y}\,]=0 \notag \\
 &\imp P(\calg)\,R_2\big( \tfrac{\pi}{3} \big)\,P(\calg^{\perp}) = P(\calg)\,R_1\big( \tfrac{\pi}{3} \big)\,P(\calg^{\perp}) =0 \notag \\
 &\imp P(\calg^{\perp}) \,\W_6(\text{-}1)\,P(\calg^{\perp}) =  \bigg(\;P(\calg^{\perp}) \,R_2\big( \tfrac{\pi}{3} \big)\,P(\calg^{\perp})\;\bigg)\notag \\
&\;\;\;\;\;\;\;\;\;\;\;\;\;\;\;\;\;\;\;\;\; \times \bigg(\;P(\calg^{\perp})\,R_1\big( \tfrac{\pi}{3} \big)\,P(\calg^{\perp})\;\bigg).
\end{align}
Alternatively stated, $\calg^{\perp}$ is closed under the operations $R_2$ and $R_1$. This implies that $P(\calg^{\perp})\,R_1(\pi/3)\,P(\calg^{\perp})$ remains an equiangular rotation in spite of the projection, i.e., there exists a reduced basis where this operator is represented by exp$(iJ\pi/3) \in SO(\text{dim}[\calg^{\perp}]\,)$, for a generator that satisfies $J=\dg{J}$, $J^2=I$ and $J=-J^t$. Similarly, $P(\calg^{\perp})\,R_2(\pi/3)\,P(\calg^{\perp})$ is also an equiangular rotation in this reduced basis. A product of two equiangular rotations has the form (\ref{specialgauge}) in a suitably chosen basis for ${\cal G}^{\perp}$: 
\begin{align} \label{ourcase}
& P(\calg^{\perp}) \,\W_6(\text{-}1)\,P(\calg^{\perp}) \notag \\
& \;\sim\;  B^t \;e^{- iS_{\perp} \pi/3}\;e^{i(1-\gamma)\,M^{\perp}_{1,2}\,\pi/3}\;B\;e^{iS_{\perp} \pi/3}, 
\end{align}
where $B \in SO(\,\text{dim}[\calg^{\perp}]\,)$.
$S_{\perp}$ and $M_{a,b}^{\perp}$ are to be distinguished from $S$ and $M_{a,b}$, which appear in the earlier discussion. Though they share similar definitions in (\ref{defineS}) and (\ref{defineMab}), they are defined on different bases: $S_{\perp}$ and $M_{a,b}^{\perp}$ are linear operators in $\calg^{\perp}$, while  $S$ and $M_{a,b}$ are linear operators in $\calg \,\cup\, \calg^{\perp}$. The form (\ref{specialgauge}) is general and does not specify whether $\gamma=+1$ or $-1$. Now we prove that our case (\ref{ourcase}) corresponds to $\gamma=+1$. Suppose $\gamma=-1$, then we may apply Lemma 1 in the reduced subspace $\calg^{\perp}$. We conclude that $P(\calg^{\perp}) \,\W_6(\text{-}1)\, P(\calg^{\perp})$ has at least the eigenvalues $\{1,1,e^{i2\pi/3},e^{\minus i2\pi/3}\}$. We have shown in App.\ref{geometricargument} that the eigenvalues $\{\text{exp}(i\vartheta)\}$ of ${\cal G}^{\perp}$ satisfy $0< |\vartheta| <2\pi/3$, if $\vartheta$ is defined on the branch $-\pi < \vartheta \leq \pi$. Thus we arrive at a contradiction. What remains is $\gamma = +1$. We have solved the eigenspectrum for such a case in App.\ref{4ntrivial}: we obtain $\text{dim}[\calg^{\perp}]/4$ sets of complex-conjugate quartets: $\{\lambda_i,\lambda_i,\lambda_i^*,\lambda_i^*\}$, as desired. \\


Combining Lemmas 1 and 2, the full spectrum is
\begin{align}
\text{eig}[\;\W_6(\text{-}1)\;] = \{\;&e^{i2\pi/3},\,e^{-i2\pi/3},\,1,\,1,\,\lambda_1,\lambda_1,\lambda_1^*,\lambda_1^*, \;\ldots,\notag \\
 &\lambda_{m-1},\lambda_{m-1},\lambda_{m-1}^*,\lambda_{m-1}^*\;\}.
\end{align}
This completes the proof of Tab. \ref{C4spectrumtable}.

\section{Alternative expression of the Wilson-loop index $\Gamma_n(\bar{k}_z)$} \label{app:equivalencepfaffian}

The weak index was first formulated in Ref.\ \onlinecite{fu2011} as an invariant involving the Pfaffian of a matrix. In this Section we show that the Pfaffian formulation is equivalent to the Wilson-loop index $\Gamma_4$ for the $C_4+T$ insulator. $\Gamma_6$ is also equivalent to a Pfaffian invariant, though the proposed formula in Ref.\ \onlinecite{fu2011} requires a clarification.

\subsection{The $C_6+T$ insulator} \label{c6equivalenceproof}

Let $V_{\sma{\pi/3}}(\boldsymbol{k})$ be the matrix representation of $\cst$ in the basis of occupied doublet bands, as defined in (\ref{sewingmatrix}). Let $\boldsymbol{k}_s$ denote momenta which are invariant under $\cst$, i.e., $-R_{2\pi/3}\boldsymbol{k}_s = \boldsymbol{k}_s$ up to a reciprocal lattice vector. These include the momenta $\Gamma,K,A,$ and $H$ of Fig.\ \ref{fig:BZ}(b). It should be noted that $V_{\sma{\pi/3}}(\boldsymbol{k}_s)$ is not skew-symmetric, thus the proposed formula (Eq. (6) in Ref.\ \onlinecite{fu2011}) for the $C_4+T$ case cannot immediately be generalized to $C_6+T$. Instead, we define $V_{\sma{\pi/3}}^a(\boldsymbol{k}_s) =  \big(\,V_{\sma{\pi/3}}(\boldsymbol{k}_s) - V_{\sma{\pi/3}}(\boldsymbol{k}_s)^t\,\big)/2$ as the skew-symmetric part of $V_{\sma{\pi/3}}$. The Wilson-loop index in the $k_z=0$ plane is alternatively expressed as
\begin{align} \label{gamma6alt}
\Gamma_6(0)   = e^{\int^{\Gamma}_K \,\text{Tr}[\,\boldsymbol{A}(\boldsymbol{k})\,] \cdot d \boldsymbol{k}}\;\frac{\text{Pf}[\,V_{\sma{\pi/3}}^a(\Gamma)\,]}{\text{Pf}[\,V_{\sma{\pi/3}}^a(K)\,]},
\end{align}
where $\text{Pf}$ denotes the Pfaffian, 
\begin{align}
\text{Tr}[\, \boldsymbol{A}(\boldsymbol{k})\,] = \sum_{i=1}^{\noc} \bra{u_{i,\boldsymbol{k}}}\,\nabla_{\boldsymbol{k}}\,\ket{u_{i,\boldsymbol{k}}}
\end{align}
is the Abelian Berry connection, and the integral follows an arbitrary path that connects $K$ to $\Gamma$. Under gauge transformations: $\ket{u_{i,\boldsymbol{k}}} \rightarrow \sum_{j}\ket{u_{j,\boldsymbol{k}}}\,\calu(\boldsymbol{k})_{ji}$ for $\calu(\boldsymbol{k}) \in U(\noc)$, $V^a \rightarrow \dg{\calu(\boldsymbol{k}_s)} \,V^a\,\calu(\boldsymbol{k}_s)^*$ which is manifestly skew-symmetric. Alternatively stated, the symmetric and anti-symmetric components of $V_{\pi/3}(\boldsymbol{k}_s)$ do not mix under gauge transformations. It follows that the entire expression on the right-hand-side (RHS) of (\ref{gamma6alt}) is gauge-invariant\cite{fu2011}.\\

\noindent \emph{Proof of (\ref{gamma6alt})}: Let us pick the shortest path that connects $K$ to $\Gamma$, in the positive quadrant of the $k_z=0$ plane. Then we relate
\begin{align}
e^{\int^{\Gamma}_K \,\text{Tr}[\,\boldsymbol{A}(\boldsymbol{k})\,] \cdot d \boldsymbol{k}} = \text{det}[\,\W_{\sma{ d \rightarrow \bar{c} }}\,],
\end{align}
where $\W_{\sma{ d \rightarrow \bar{c} }}$ is a Wilson line defined in App.\ref{app:rvWilson}. In the real, periodic basis of App.\ \ref{existrealgauge}, $\W_{\sma{ d \rightarrow \bar{c} }} \in O(\noc)$ and the matrix $V_{\pi/3}$ is chosen to have the canonical form (\ref{canonicalform}), thus
\begin{align}
V_{\sma{\pi/3}}^a(\boldsymbol{k}_s) = \si(\pi/3) \,E(\boldsymbol{k}_s)\,i \,S\,E(\boldsymbol{k}_s)^t,
\end{align}
where $E(\boldsymbol{k}_s) \in O(\noc)$ and $S$ is defined in (\ref{defineS}). Exploiting the identity Pf$[ \,E(\boldsymbol{k}_s)\,iS\,E(\boldsymbol{k}_s)^t\,]$= det$[\, E(\boldsymbol{k}_s)\,]$, we derive 
\begin{align}
&e^{\int^{\Gamma}_K \,\text{Tr}[\,\boldsymbol{A}(\boldsymbol{k})\,] \cdot d \boldsymbol{k}}\;\frac{\text{Pf}[\,V_{\sma{\pi/3}}^a(\Gamma)\,]}{\text{Pf}[\,V_{\sma{\pi/3}}^a(K)\,]} \notag \\
\eq \text{det}[\,E(d)\,\W_{\sma{ d \rightarrow \bar{c} }}\,E(c)^t\,] \in \{1,\mo\}.
\end{align}
In App.\ \ref{appsec:c6proof}, we identify $\Gamma_6(0)=\text{det}[\,E(d)\,\W_{\sma{ d \rightarrow \bar{c} }}\,E(c)^t\,]$.

\subsection{The $C_4+T$  insulator}

Let $V_{\sma{\pi/2}}(\boldsymbol{k})$ be the matrix representation of $\cft$ in the basis of occupied doublet bands, as defined in (\ref{sewingmatrix}). Let $\boldsymbol{k}_s$ denote momenta which are invariant under $\cft$, i.e., $-R_{\pi/2}\boldsymbol{k}_s = \boldsymbol{k}_s$ up to a reciprocal lattice vector; these include the momenta $\Gamma,M,Z,$ and $A$ of Fig.\ \ref{fig:BZ}(a). It should be noted that $V_{\sma{\pi/2}}(\boldsymbol{k}_s)$ is skew-symmetric\cite{fu2011}.  The Wilson-loop index in the $k_z=0$ plane is alternatively expressed as
\begin{align} \label{gamma4alt}
\Gamma_4(0)   = e^{\int^{\Gamma}_M \,\,\text{Tr}[\,\boldsymbol{A}(\boldsymbol{k})\,] \cdot d \boldsymbol{k}}\;\frac{\text{Pf}[\,V_{\sma{\pi/2}}(\Gamma)\,]}{\text{Pf}[\,V_{\sma{\pi/2}}(M)\,]},
\end{align}
where the integral follows an arbitrary path that connects $M$ to $\Gamma$. The RHS of (\ref{gamma4alt}) is gauge-invariant\cite{fu2011}. \\

\noindent \emph{Proof of (\ref{gamma4alt})}: The steps are closely analogous to the $C_6+T$ case. Let us pick the shortest path that connects $M$ to $\Gamma$, in the positive quadrant of the $k_z=0$ plane. Then we relate
\begin{align}
e^{\int^{\Gamma}_M \,\,\text{Tr}[\,\boldsymbol{A}(\boldsymbol{k})\,] \cdot d \boldsymbol{k}} = \text{det}[\,\W_{\sma{b \nearrow \bar{a}}}\,],
\end{align}
where $\W_{\sma{b \nearrow \bar{a}}}$ is a Wilson line defined in App.\ref{app:rvWilson}. A basis is found where $V_{\pi/2}$ has the canonical form (\ref{canonicalform}), or equivalently,
\begin{align}
V_{\sma{\pi/2}}(\boldsymbol{k}_s)=E(\boldsymbol{k}_s)\,i\,S\,E(\boldsymbol{k}_s)^t; \;\;\; E(\boldsymbol{k}_s) \in O(\noc).
\end{align}
It follows that
\begin{align}
 &e^{\int^{\Gamma}_M \,\,\text{Tr}[\,\boldsymbol{A}(\boldsymbol{k})\,] \cdot d \boldsymbol{k}}\;\frac{\text{Pf}[\,V_{\sma{\pi/2}}(\Gamma)\,]}{\text{Pf}[\,V_{\sma{\pi/2}}(M)\,]}  \notag \\
\eq \text{det}[\,E(b)\,\W_{\sma{b \nearrow \bar{a}}}\,E(a)^t\,] \in \{1,\mo\}.
\end{align} 
In App.\ \ref{appsec:c4proof} , we identify 
\begin{align}
\Gamma_4(0)=\text{det}[\,E(b)\,\W_{\sma{b \nearrow \bar{a}}}\,E(a)^t\,].
\end{align}

\section{Choice of spatial origin for the bent Wilson loop} \label{app:origin}

As is well-known in the geometric theory of polarization\cite{atala2013}, translating the spatial origin multiplies each Berry phase by a global $U(1)$ phase. This $U(1)$ variance applies to any non-contractible Wilson loop in the 3D BZ, including the bent loops that we are proposing, as we now show. Consider a non-contractible loop ($l$), with start point $\boldsymbol{k^{\sma{(0)}}}$ and end point $\boldsymbol{k^{\sma{(0)}}}+\boldsymbol{G}$, with $\boldsymbol{G}$ a reciprocal lattice vector. The Berry-phase spectrum for this loop comprises the unimodular eigenvalues of the operator
\begin{align} 
\hat{\W}[l] = \cald(\boldsymbol{G}) \; {\displaystyle \prod_{\alpha}^{\boldsymbol{k^{\sma{(0)}}}+\boldsymbol{G} \leftarrow \boldsymbol{k^{\sma{(0)}}}}}\, P(\boldsymbol{k^{\sma{(\alpha)}}}),
\end{align}
as deducible from Eq.\ (\ref{eq:wilsonprojection}) and the periodic gauge: $\ket{u_{i,\boldsymbol{k}+\boldsymbol{G}}} = \cald(\boldsymbol{G})^{\mo}\,\ket{u_{i,\boldsymbol{k}}}$; recall the definition of  $[\cald(\boldsymbol{G})]_{\ab}$ as $\delta_{\ab}\,e^{i\boldsymbol{G}\cdot \boldsymbol{r_{\alpha}}}$ in App.\ \ref{app:rvWilson}, where $ \boldsymbol{r_{\alpha}}$ denotes the spatial embedding of an orbital (labeled by $\alpha$) within the unit cell. The effect of translating our spatial origin by $\boldsymbol{\delta}$ is that $\hat{\W}$ is translated by a global $U(1)$ phase: $\hat{\W} \rightarrow e^{i\boldsymbol{G}\cdot\boldsymbol{\delta}}\hat{\W}$. This follows from two observations: (i) the tight-binding Hamiltonian\ (\ref{tbHam}) is invariant under this translation, hence the projections $P(\boldsymbol{k})$ likewise. (ii) The spatial embedding of the orbitals are modified as $ \boldsymbol{r_{\alpha}} \rightarrow  \boldsymbol{r_{\alpha}} +\boldsymbol{\delta} $ for each $\alpha$, thus $\cald(\boldsymbol{G}) \rightarrow \cald(\boldsymbol{G})e^{i\boldsymbol{G}\cdot\boldsymbol{\delta}}$. \\

This $U(1)$ variance has implications for the bent Wilson loop $\hat{W}[l_4(\bar{k}_z)]$, for $\bar{k}_z \in \{0,\pi\}$ and the loop $l_4(\bar{k}_z)$ defined in Sec.\ \ref{sec:cnt}. For example, $l_4(0)$ is the loop connecting $M-\Gamma-M$ in the $k_z=0$ plane, as depicted in red in Fig.\ \ref{fig:BZsWloops}(a); to be explicit, we may choose coordinates such that the loop starts at $\boldsymbol{k^{\sma{(0)}}}=M$ and ends at $\boldsymbol{k^{\sma{(0)}}}+\boldsymbol{G} = M + 2\pi \hat{x}/a$, with $a$ a lattice constant in the tetragonal lattice. As an example, we consider a tetragonal lattice composed of two interpenetrating sublattices, which are correspondingly colored red and blue in Fig.\ \ref{fig:tetragonal}(a). We label our orbital basis by $ \alpha =1$ (resp.\ $2$) for $p_x+ ip_y$ (resp. $p_x- ip_y$) orbitals on the blue sublattice, and $ \alpha =3$ (resp.\ $4$) for $p_x- ip_y$ (resp. $p_x+ ip_y$) orbitals on the red sublattice. Suppose the spatial origin lies on an atomic site in the blue sublattice, e.g., the corner of the cube in Fig.\ \ref{fig:tetragonal}(a).  With this choice of origin, the spatial embeddings are $\boldsymbol{r}_1=\boldsymbol{r}_2=0$, and $\boldsymbol{r}_3=\boldsymbol{r}_4 = \Delta \hat{z}$ corresponds to the vertical offset between the two sublattices -- clearly, with $G=2\pi \hat{x}/a$, the phase factor $e^{i\boldsymbol{G}\cdot \boldsymbol{r_{\alpha}}}$ is trivially unity for all $\alpha$. In Sec.\ \ref{sec:cnt}, we defined the trivial (resp. weak) phase by their Berry-Zak spectrum: $\W[l_4(0)]=\W[l_4(\pi)]={+I}$ (resp. ${-I}$). We point out that this definition is only meaningful if we specify the spatial origin, which in this case lies on an atomic site. There exists a second choice of origin that manifests the four-fold rotational symmetry, e.g., the cross in Fig.\ \ref{fig:tetragonal}(a) and (b). With this second choice, $\boldsymbol{r}_1=\boldsymbol{r}_2=a(\hat{x}+\hat{y})/2$ (illustrated by the green arrow in Fig.\ \ref{fig:tetragonal}(b)), and $\boldsymbol{r}_3=\boldsymbol{r}_4=a(\hat{x}+\hat{y})/2 + \Delta \hat{z}$. Now the phase factor $e^{i\boldsymbol{G}\cdot \boldsymbol{r_{\alpha}}}={-1}$ for all $\alpha$, and the \emph{same} trivial (resp. weak) phase is characterized by $\W[l_4(0)]=\W[l_4(\pi)]={-I}$ (resp. ${+I}$).

\bibliography{bib_13apr}

\begin{thebibliography}{54}
\bibitem [{foo()}]{footnote1}%
  \BibitemOpen
  \href@noop {} {}\bibinfo {note} {In principle, we may consider a second
  family of $C_6$ Wilson loop which connects $K_2-\Gamma-K_2$, where $K_2$ is
  the second $C_3$-invariant point on the vertex of the hexagonal BZ; due to
  $C_6$ symmetry, this loop is related to (\ref{c6wl}) by a unitary
  transformation, and thus contains no new information.}\BibitemShut {Stop}%
\end{thebibliography}

\end{document}